\documentclass{article}

\usepackage[english]{babel}

\usepackage[letterpaper,top=2cm,bottom=2cm,left=3cm,right=3cm,marginparwidth=1.75cm]{geometry}

\usepackage{url}
\usepackage{bm}
\usepackage{array}
\usepackage{amsmath}
\usepackage{amssymb}
\usepackage{graphicx}
\usepackage{subcaption}
\usepackage{multirow}
\usepackage{booktabs}
\usepackage[colorlinks=true, allcolors=blue]{hyperref}
\usepackage{makecell}
\usepackage[table]{xcolor}

\title{Prediction of Viscoelastic Droplet Impact Dynamics Using a Vision Transformer-Based Approach}
\author{
\begin{tabular}{c c}
Diego Alecsander de Aguiar$^{1}$ &
Cassio Machiaveli Oishi$^{1}$ \\[0.3em]
\texttt{diego.aguiar@unesp.br} &
\texttt{cassio.oishi@unesp.br}
\end{tabular}
\\[1em]
$^{1}$Departamento de Matemática e Computação, Faculdade de Ciências e Tecnologia,\\
Universidade Estadual Paulista ``Júlio de Mesquita Filho'',\\
Presidente Prudente, Brazil
}
\date{}

\begin{document}
\maketitle

\begin{abstract}

Droplet impact on solid surfaces is a complex fluid dynamics problem with different applications in processes such as spray cooling, inkjet printing and pharmaceutical processing. Although numerical simulations are widely used to investigate these dynamics, their computational cost becomes significant when multiple parametric variations are explored. In this work, we investigate the use of a Video Vision Transformer (ViViT) architecture to predict the temporal evolution of viscoelastic droplets impacting solid surfaces using volume fraction fields from the Volume of Fluids (VOF) interface representation method. In Newtonian fluids, the impact dynamics are mainly characterized by the Reynolds number $Re$, which represents the ratio of inertial forces to viscous forces, and the Weber number $We$, which represents the ratio of inertial forces to surface tension forces. For viscoelastic fluids, however, additional parameters are required to account for elastic effects, namely the solvent viscosity ratio $\beta$ and the Weissenberg number $Wi$, increasing both the complexity and the computational cost of numerical simulations. Instead of numerically simulating the entire droplet dynamics, the proposed deep learning approach uses only the initial 10\% to 20\% of the simulation to predict the remaining evolution of the system. Depending on the prediction configuration, this strategy reduces the computational cost by approximately 80\% to 90\% compared to a full numerical simulation. The ViViT maintains physically consistent predictions across different parameters and prediction horizons, successfully capturing both spreading and bouncing regimes while preserving geometric properties and structural similarity of the droplet dynamics. Since volume fraction fields can also be extracted from videos of real experiments, the proposed framework could potentially be adapted to incorporate experimental data during training, further improving the physical fidelity of the predicted dynamics.

\vspace{0.2cm}


\end{abstract}

\section{Introduction}

The dynamics of droplet impact, characterized by complex behaviors, play an important role in a wide range of industrial and scientific applications such as spray cooling, food and pharmaceutical processing, criminal forensics, inkjet printing and 3D organ bioprinting \cite{Gao2025, Andrade2013, Bolleddula2010, Smith2018, Detlef2022, Sanjairaj2018}. Studies on droplet impact phenomena have been extensively conducted for Newtonian fluids \cite{Rein1993, Christophe2004, Jae2016} and non-Newtonian fluids \cite{Li2009, Oishi2019, Cassio2019, Kindness2024}. These studies explore different key aspects such as the relevant dimensionless numbers and their influence on maximum droplet spreading, as well as the different regimes before and after impact, including spreading, bouncing, and splashing.

A deeper understanding of droplet impact on solid surfaces has been achieved through the combined efforts of analytical (theoretical foundations of governing equations and principles) \cite{Siddhartha2007, Ilia2009}, experimental (validation of theoretical assumptions) \cite{Bertola2009, Nick2014, Jae2016, Shiji2018} and numerical (approximations of physical behaviors) \cite{Lorstad2004, Oishi2012, Hua2015, Zhuan2022} methodologies. These advances play an important role not only in validating, but also in improving the fidelity of computational simulations.

While the impact dynamics of Newtonian droplets are mainly characterized by the Reynolds number $Re$, which represents the balance between inertial and viscous forces, and the Weber number $We$, which represents the balance between inertial and surface tension forces, the behavior of viscoelastic droplets is considerably more complex. In addition to these dimensionless numbers, viscoelastic fluids introduce elastic effects that are characterized by the solvent viscosity ratio $\beta$ and the Weissenberg number $Wi$. The interaction between inertia, viscosity, surface tension and elasticity can significantly affect droplet deformation and transition between impact regimes \cite{Crooks2000, Bertola2009, Jung2013, Izbassarov2016, Wang2017, Rostami2024}, such as spreading and bouncing. Furthermore, exploring a larger parametric space becomes significantly more challenging and computationally expensive.

Scientific machine learning has emerged as an important tool for improving and analyzing computational fluid dynamics (CFD) simulations \cite{Jingzu2023, Hossein2023, Oishi2024, Omata2019}, especially in scenarios where traditional numerical methods are too computationally expensive, slow or limited in accuracy. For example, \cite{Jiguo2023} developed a universal model to predict the maximum spreading of a droplet impacting a smooth surface using physical parameters and non-dimensional numbers, eliminating the need for a full simulation. Similarly, \cite{Yu2026} proposed a model to predict the outcome of binary droplet collision (coalescence, bouncing or separation) that addresses multiple parametric variations such as Weber number $We$, impact parameter $B$, size ratio $\Delta$ and Ohnesorge number $Oh$. Furthermore, \cite{Vulf2025} developed a model for classifying the collision of suspension droplets against solid dry surfaces (splashing, breaking up, and rebound) with varied particle properties, liquid parameters, substrate properties, droplet velocity and substrate inclination. More recently, \cite{Dreisbach2025} introduced a physics-informed neural network framework that combines fluid dynamics equations with visual observations to reconstruct 3D velocity, pressure, and fluid phases from specific moments of experimental droplet impact images.

In particular, transformer-based architectures \cite{Vaswani2017} have recently gained significant attention in fluid dynamics due to their strong ability to capture both spatial and temporal long-range dependencies in data \cite{Yu2023, Solera2024, Xu2024, Khan2025, Wang2026}, with several works extending these models to handle image-based data through the Vision Transformer architecture \cite{Alexey2021}. For example, \cite{Kazemi2025} proposed a deep learning model based on a U-Net with Vision Transformer and time embedding applied to subsurface flow modeling in reservoir simulation to predict saturation and pressure fields from permeability maps. Similarly, in the field of reservoir characterization, \cite{Temizel2025} uses ViTs for permeability prediction on diverse rock samples, surpassing traditional Convolutional Neural Networks (CNNs). \cite{Kang2023} proposed the CFDformer, a surrogate model combining a ViT and a U-shaped CNN to predict fluid flow fields (velocity and pressure) from geometry information and boundary conditions, outperforming baseline models. Using data from CFD simulations of an argon jet injected into a quiescent nitrogen environment for training, \cite{Yalamanchi2026} uses a hierarchical Vision Transformer (SwinV2-UNet) for autoregressive spatiotemporal rollouts to forecast flow evolution (predicting step $t+1$ from step $t$), as well as for feature transformations (reconstructing unobserved velocity fields from density fields).

Specifically in the field of droplet dynamics, however, such approaches remain largely underexplored. Furthermore, in the context of prediction modeling, most existing approaches focus on single-step or independent inputs to predict future states, without capturing a broader temporal context. Although several deep learning approaches for video-based prediction with spatial-temporal modeling have been proposed in other areas, including recurrent architectures \cite{Oliu2018}, 3D Convolutional Neural Networks (3D CNNs) \cite{Tran2015} and Convolutional LSTMs (ConvLSTMs) \cite{Shi2015}, their application to droplet dynamics remains limited. This is mainly because many of these methods rely on local spatial patterns and relatively short temporal context windows, which becomes especially challenging for problems involving interface deformation and regime transitions over time.

To address this limitation, this work explores the use of a Video Vision Transformer (ViViT) \cite{Arnab2021}, an extension of the Vision Transformer designed to process spatio-temporal data, for the prediction of droplet impact dynamics on solid surfaces. Unlike convolutional architectures, which analyze the data region by region using filters that gradually combine information across layers, the ViViT is naturally suited to capture global dependencies between elements in a sequence and maintain long-range temporal context through the attention mechanism. This reduces the loss of information commonly observed in recurrent and convolution-based approaches when dealing with longer prediction horizons. This characteristic is particularly important for reliably capturing complex droplet dynamics, such as spreading and bouncing, while still enabling significant reductions in computational cost compared to full numerical simulations.

The dataset used in this study consists of volume fractions obtained from numerical simulations of viscoelastic droplets impacting a solid surface, considering variations in the Reynolds number $Re$, Weber number $We$, solvent viscosity ratio $\beta$ and Weissenberg number $Wi$. The simulations are run using the open-source flow solver Basilisk \cite{Popinet2013}, which uses the Volume of fluid (VOF) surface tracking technique. To reduce computational costs, an axisymmetric formulation is adopted. In our proposed framework, the predicted quantities correspond to the temporal evolution of the volume fraction fields, which describe the droplet morphology over time. Since these interface representations can also be obtained from segmented frames of experimental videos, the proposed approach could potentially incorporate experimental data during training, improving the physical fidelity of the learned dynamics.

To assess the advantages of our architecture, its performance is compared with that of a simpler baseline model, namely a Multilayer Perceptron (MLP). This comparison highlights the limitations of models that lack explicit mechanisms to capture temporal dependencies. Additionally, the adoption of a ViViT-based approach was motivated by its potential to preserve physical consistency, ensuring a more faithful representation of the underlying dynamics associated with each set of droplet parameters. Three input-output configurations were evaluated: 50 time steps to predict one step, 50 steps to predict the subsequent 50 steps, and 100 steps to predict the subsequent 100 steps. Model performance is evaluated under four aspects: predictive performance, image quality, preservation of geometric properties, and computational cost.

\section{Methods}

\begin{figure}[htpb]
    \centering
    \includegraphics[width=1.0\linewidth]{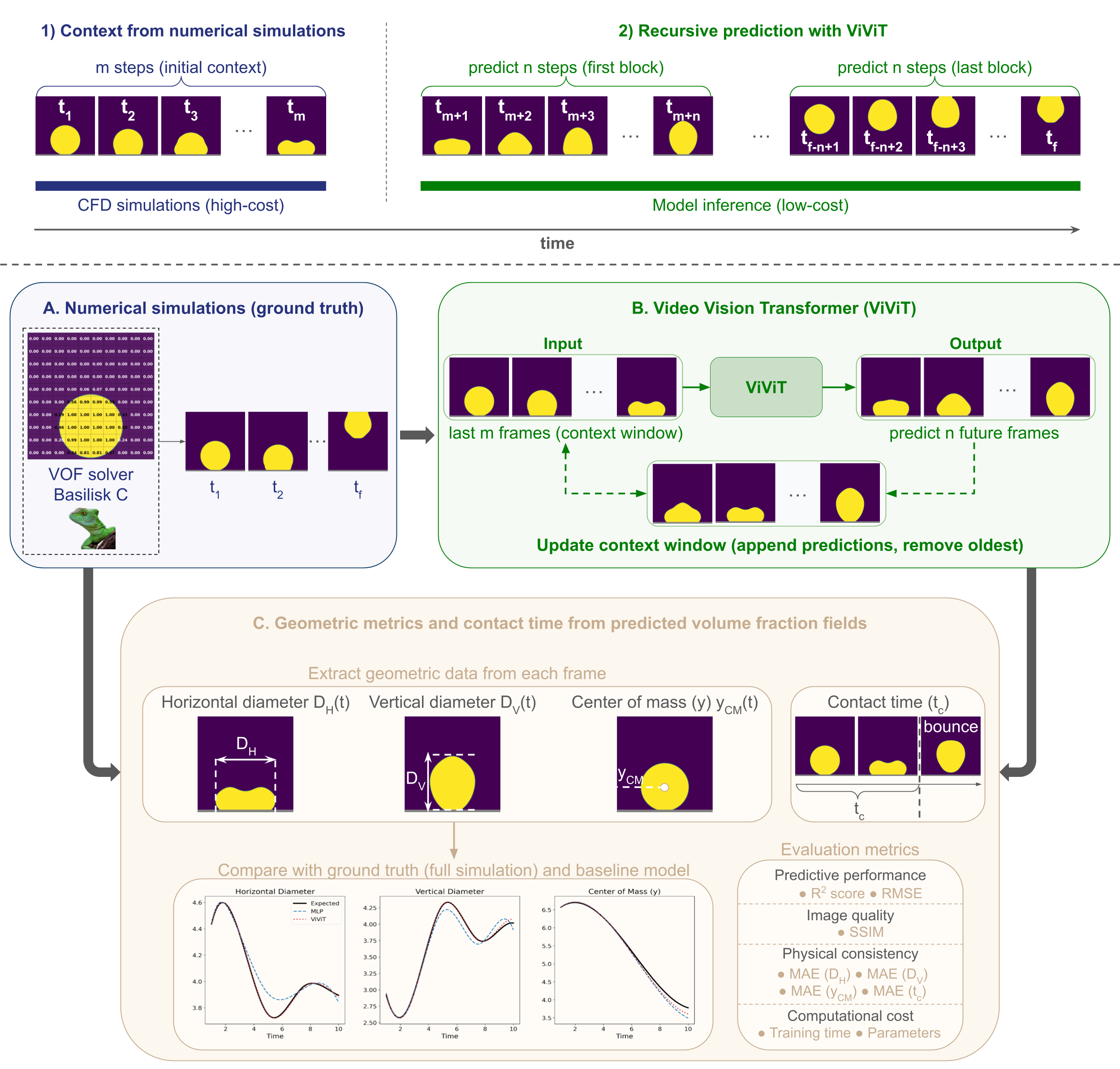}
    \caption{General pipeline of the proposed framework. (1) Initial temporal context obtained from CFD simulations (high-cost) is used as input. (2) ViViT framework (low-cost) recursively predicts future steps. (A) Volume fraction fields generated from Basilisk C numerical simulations are used as ground truth data. (B) A Video Vision Transformer (ViViT) receives the last $m$ frames as temporal context and recursively predicts the next $n$ future frames, updating the context window after each prediction block. (C) Geometric properties and contact time are extracted from the predicted volume fraction fields and compared with the reference simulations, evaluating predictive performance, image structure quality, physical consistency and computational cost.}
    \label{fig:1_pipeline}
\end{figure}

Droplet impact on solid surfaces exhibits complex behaviors, such as spreading and bouncing, which are highly sensitive to initial conditions. In this section, we present a machine learning approach trained using data from numerical simulations to predict the temporal evolution of viscoelastic droplet impacts. We compare a simple Multilayer Perceptron with a Vision Transformer architecture adapted for video input, based on the implementation proposed by \cite{Arnab2021}. Using an initial context of the droplet impact, the models recursively predict future volume fraction fields, allowing the subsequent dynamics, including spreading and bouncing behaviors, to be generated without performing the full numerical simulation. A general overview of the proposed prediction pipeline is illustrated in Fig. \ref{fig:1_pipeline}, where only 10\% or 20\% of the trajectory is generated by high-cost CFD simulations and the remaining 80\% or 90\% is predicted through low-cost model inference. To facilitate reproducibility and future developments, the complete implementation developed in this work, including a graphical user interface (GUI), has been made publicly available and can be accessed through the repository listed in Section \ref{sec:code}.

\subsection{Dataset}

The dataset used in this study is composed of images derived from numerical simulations of viscoelastic droplets impacting a solid surface using the open-source flow solver Basilisk \cite{Popinet2013}, which solves the two-phase incompressible axisymmetric Navier-Stokes equations using the Volume of fluid (VOF) surface tracking technique. More specifically, the volume fractions at each time step are used as input to the models.

\subsubsection{Mathematical modeling}

The governing equations are the continuity and momentum equations:

\begin{align}
    & \nabla \cdot \textbf{u} = 0, \\
    & \rho \left( \frac{\partial \textbf{u}}{\partial t} + \nabla \cdot (\textbf{uu}) \right) = - \nabla p + \nabla \cdot \boldsymbol{\tau} + \textbf{f}_s + \rho \textbf{g},
\end{align}
where $\mathbf{u}$ and p are the velocity and pressure fields, $\rho$ is the fluid density, $\textbf{f}_s$ is the surface tension force, $\textbf{g}$ is gravity and $\boldsymbol{\tau}$ is the extra-stress tensor composed of solvent and polymeric contributions:

\begin{equation}
    \boldsymbol{\tau} = 2 \eta_s \mathbf{D} + \boldsymbol{\tau}_p,
\end{equation}
in which $\eta_s$ is the solvent viscosity, $D = \frac{1}{2} \left ( \nabla \mathbf{u} + \nabla \mathbf{u}^T \right )$ is the strain-rate tensor, and $\boldsymbol{\tau}_p$ is the polymeric stress, which represents the non-Newtonian contribution in the flow. The polymeric stress evolution follows the log-conformation formulation of the Oldroyd-B constitutive model \cite{Raanan2005}:

\begin{equation}
  \frac{\partial \boldsymbol{\Psi}}{\partial t} +(\mathbf{u}\cdot\nabla)\boldsymbol{\Psi} - (\boldsymbol{\Omega}\boldsymbol{\Psi} - \boldsymbol{\Psi}\boldsymbol{\Omega})
  - 2 \mathbf{D} = -\frac{1}{Wi}\left(e^{-\boldsymbol{\Psi}} - \mathbf{I}\right),
\end{equation}
where $\boldsymbol{\Psi} = \ln(\mathbf{A})$ is the logarithm of the conformation tensor $\mathbf{A}$, $\boldsymbol{\Omega}$ is the vorticity tensor, and $Wi$ is the Weissenberg number. The polymeric stress is recovered as

\begin{equation}
  \boldsymbol{\tau}_p = \frac{1 - \beta}{Wi} \left( e^{\boldsymbol{\Psi}} - \mathbf{I} \right),
\end{equation}
where $\eta_p$ is the polymer viscosity contribution and $\beta = \frac{\eta_s}{\eta_s + \eta_p}$ is the solvent viscosity ratio.

The surface tension force $\textbf{f}_s$ is included using the Continuous Surface Force (CSF) model \cite{Jeremiah1992}:

\begin{equation}
  \mathbf{f}_s = \frac{1}{We}\, \kappa \nabla f, \qquad
  \kappa = \nabla \cdot \left( \frac{\nabla f}{|\nabla f|} \right),
\end{equation}
in which $\kappa$ is the curvature of the interface, $We$ is the Weber number, and $f$ is the volume fraction field, where $f = 1$ corresponds to the primary fluid, $f = 0$ corresponds to the secondary fluid (air) and the intermediate values $0 < f < 1$ identify the surface cells containing a ratio of both fluids.

The solver allows the problem to be configured in a dimensionless form through the use of non-dimensional numbers such as the solvent viscosity ratio $\beta$ and the Reynolds $Re$, Weber $We$ and Weissenberg $Wi$ numbers:

\begin{equation}
    Re = \frac{\rho U L}{\eta_s + \eta_p}, \hspace{0.3cm} We = \frac{\rho U^2L}{\sigma}, \hspace{0.3cm} Wi = \frac{\lambda U}{L},
\end{equation}
where $U$ and $L$ are, respectively, the characteristic velocity and length, $\sigma$ is the surface tension, and $\lambda$ is the relaxation time of the viscoelastic fluid. 

The initial condition corresponds to a droplet of radius $R$ located directly above the solid surface, surrounded by air, moving downward with velocity $U$. In our tests, $L$ is defined as the radius of the droplet. The solid surface is defined as a no-slip and impermeable region where:

\begin{equation}
\begin{cases}
    u_n = 0, & \text{(impermeability)} \\
    u_{\text{fluid}} = u_{\text{wall}}, & \text{(no-slip)}
\end{cases}
\end{equation}

Depending on the interaction dynamics, the droplets may spread through the solid surface and, in some cases, bounce after an initial contact time. 

\subsubsection{Numerical simulations}

To generate a balanced dataset with both spreading and bouncing cases, 180 numerical simulations were performed using different combinations of dimensionless numbers ($Re$, $We$, $Wi$, and $\beta$) and a fixed initial droplet radius of $R = 2$. All spatial and temporal quantities are non-dimensionalized.

Table \ref{tab:parametros_simulacoes} summarizes the parametric variations. The computational domain spans $x, y \in [0, 8]$, with the axisymmetric condition applied along the $y$-axis. A uniform grid of size $128 \times 128$ was used for all simulations, with $\Delta x = \Delta y = 0.0625$. The simulation is run until $T_{\text{max}} = 10$ with a time step of $\Delta t = 0.0001$. Data are collected every $\Delta t = 0.02$, generating a total of 500 snapshots per simulation.

\begin{table}[h!]
\centering
\caption{Summary of the numerical dataset used in this study.}
\label{tab:parametros_simulacoes}
\begin{tabular}{|cc|}
\hline
\textbf{Parameter} & \textbf{Range} \\
\hline
$Re$              & [2.5, 150] \\
$We$              & [0.5, 40] \\
$\beta$           & [0.1, 0.9] \\
$Wi$              & [1.0, 5.0] \\
\hline
\textbf{Total simulations} & 90 bouncing and 90 spreading cases were randomly sampled \\
\hline
\end{tabular}
\end{table}

From each simulation, we extract the temporal evolution of the volume fractions. Volume fraction fields were chosen for prediction because they capture the morphology of the droplet and have broad applicability, since they can also be easily obtained from experimental images. The original axisymmetric field is first cropped on the right side to retain only half of its width. Afterward, the reduced image is mirrored along the axisymmetric axis to generate a $128 \times 128$ representation.

To select the mesh resolution for this work, a mesh convergence analysis was performed considering two droplet impact cases, one in the bouncing regime and another in the spreading regime, simulated with meshes of $64 \times 64$, $128 \times 128$, $256 \times 256$ and $512 \times 512$. The objective was to identify the minimum resolution capable of preserving the expected physical behavior while minimizing computational cost.

\begin{figure}[htpb]
    \centering
    \begin{subfigure}{0.95\linewidth}
        \centering
        \includegraphics[width=1.0\linewidth]{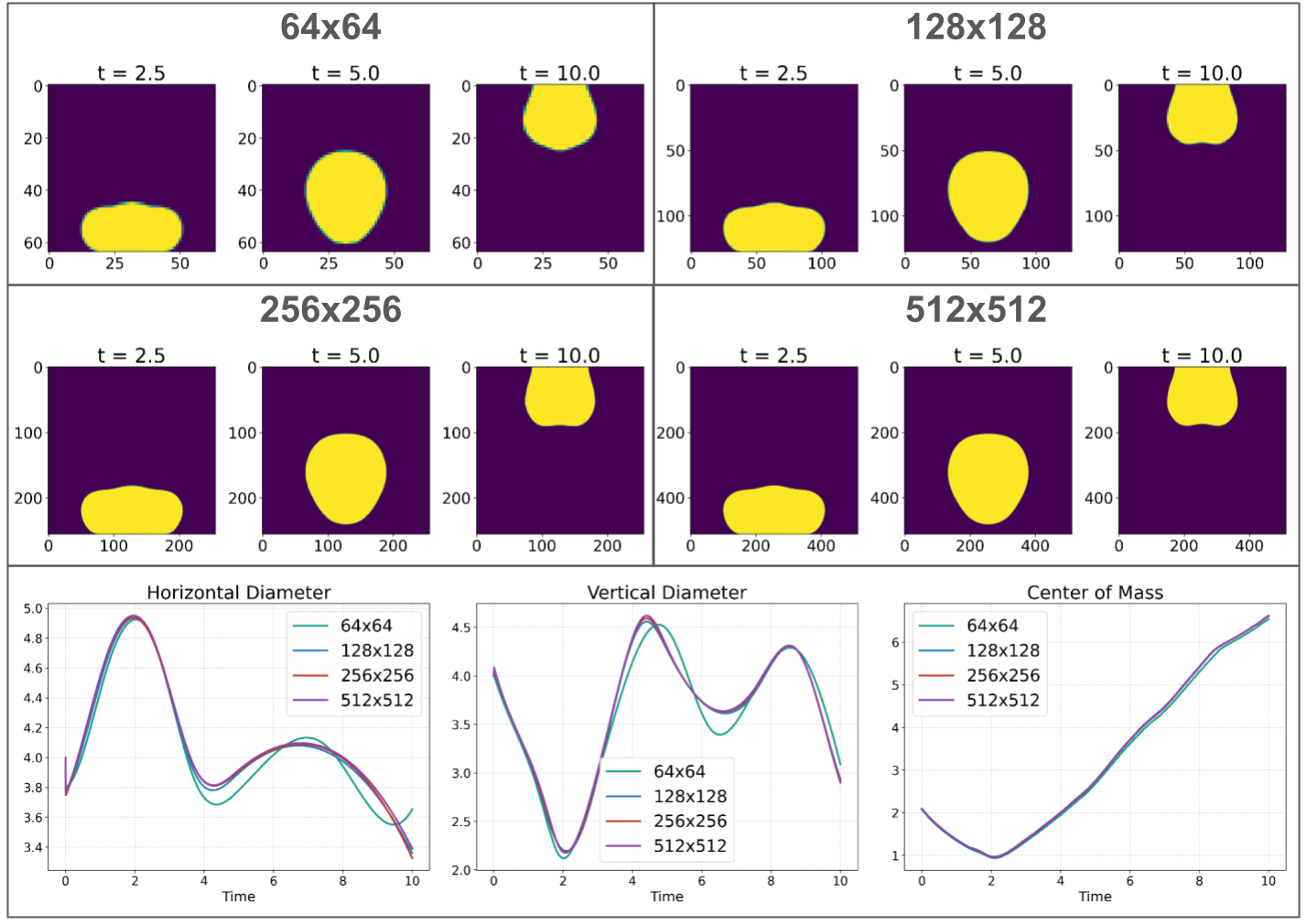}
        \caption{Bouncing case - Re = 68.06, We = 0.5, $\beta = 0.9$ and Wi = 3.0.}
        \label{fig:comparacao_malhas_bouncing}
    \end{subfigure}
    \begin{subfigure}{0.95\linewidth}
        \centering
        \includegraphics[width=1.0\linewidth]{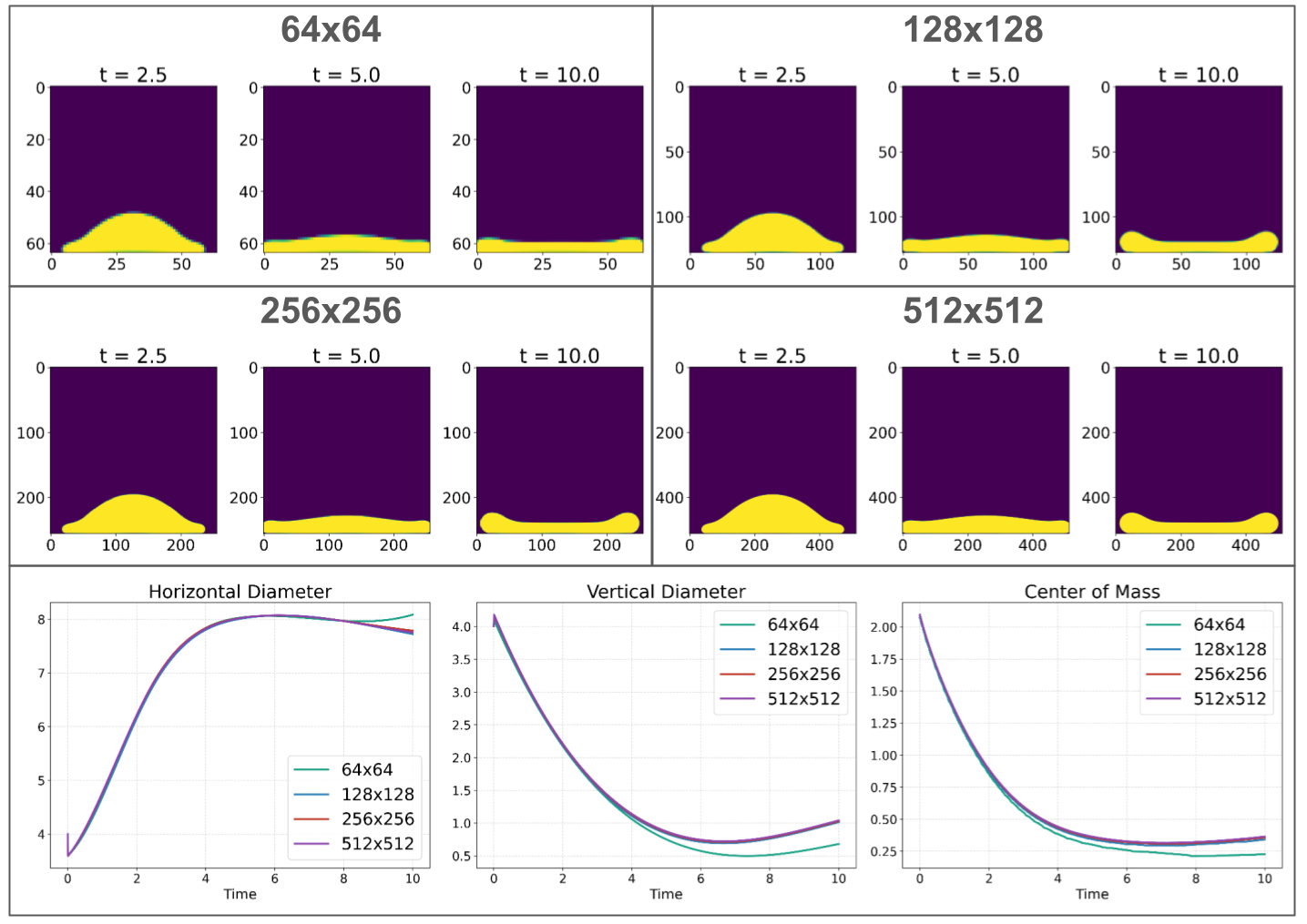}
        \caption{Spreading case - Re = 18.89, We = 35.61, $\beta = 0.9$ and Wi = 5.0.}
        \label{fig:comparacao_malhas_spreading}
    \end{subfigure}
    \caption{Comparison of the droplet impact dynamics obtained with different mesh resolutions for bouncing and spreading cases.}
    \label{fig:comparacao_malhas}
\end{figure}

Both qualitative and quantitative analyses were performed. Qualitatively, the temporal evolution of the volume fraction fields was evaluated to verify the consistency of the droplet interface representation across different mesh resolutions. Quantitatively, the horizontal diameter, vertical diameter and y-axis center of mass were extracted over time for each case. These comparisons are presented in Fig. \ref{fig:comparacao_malhas}.

Additionally, the contact times $t_c$ (time during which the droplet remained in contact with the solid surface) obtained for the bouncing case were 4.87, 4.73, 4.66 and 4.62 for the $64 \times 64, 128 \times 128, 256 \times 256$ and $512 \times 512$ meshes, respectively, while the spreading case remained in contact with the surface throughout the simulation ($t_c = 10.0$) for all mesh resolutions. Table \ref{tab:comparacao_malhas} presents the computational execution times required to run the full simulations ($t_{\text{sim}}$) for both analyzed cases, together with an estimate of the potential reduced execution time obtained using the proposed model with the largest initial context tested ($t_{\text{model}}$).

\begin{table}[h!]
\centering
\caption{Computational cost for different mesh resolutions. $t_{\text{model}} = 0.2 t_{\text{sim}} + t_{\text{inf}}$, where only 20\% of the trajectory is simulated and 80\% is predicted by the model, with negligible inference time $t_{\text{inf}}$.}
\label{tab:comparacao_malhas}
\begin{tabular}{|ccccc|}
\hline
\textbf{Mesh} & \textbf{Bouncing $t_{\text{sim}}$} & \textbf{Bouncing $t_{\text{model}}$} &
\textbf{Spreading $t_{\text{sim}}$} & \textbf{Spreading $t_{\text{model}}$} \\
\hline
64 $\times$ 64   & 0.5h & 0.1h & 0.5h & 0.1h\\
128 $\times$ 128 & 7.25h & 1.45h & 6.9h & 1.38h \\
256 $\times$ 256 & 21.38h & 4.28h & 21.95h & 4.39h \\
512 $\times$ 512 & 64.64h & 12.93h & 74.26h & 14.85h \\
\hline
\end{tabular}
\end{table}

The results indicated that the $64 \times 64$ mesh presents noticeable differences in interface, geometric properties and contact time for the bouncing sample when compared with the finer meshes. In contrast, the $128 \times 128$, $256 \times 256$ and $512 \times 512$ meshes exhibit very similar behaviors, with only negligible variations in geometric properties and contact time. Additionally, the computational cost increases significantly as the mesh is refined. Therefore, the $128 \times 128$ mesh was selected as the lowest resolution capable of preserving the physical behavior observed in the finer meshes while maintaining a lower computational cost.
    
\begin{figure}[htpb]
    \centering
    \includegraphics[width=1.0\linewidth]{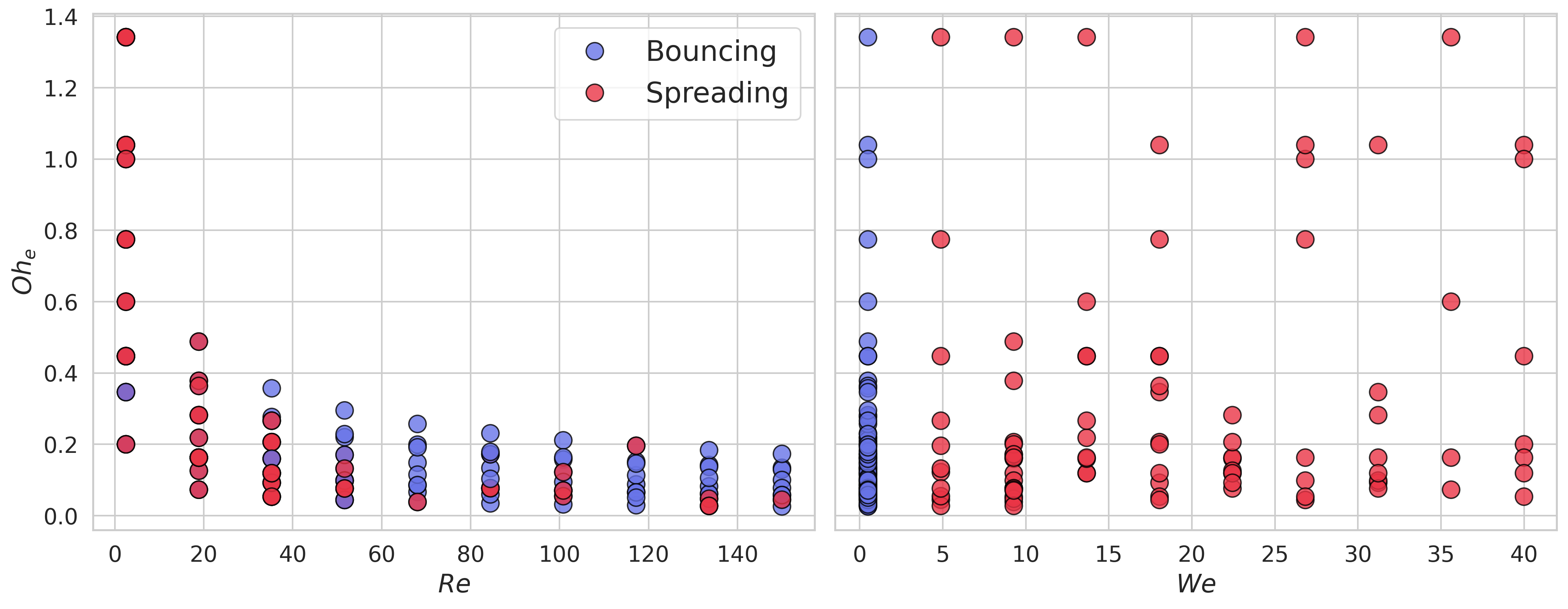}
    \caption{Distribution of bouncing and spreading cases.}
    \label{fig:1_dataset}
\end{figure}

The distribution of bouncing and spreading regimes is shown in Fig. \ref{fig:1_dataset} as a function of $Re$ and $We$ on the x-axis and the elastic Ohnesorge number $Oh_e = \frac{(1 - \beta) Wi}{Re}$) \cite{Cassio2019} on the y-axis. As observed, bouncing is mainly associated with low $We$, while spreading is distributed across a wider range of $We$ and $Oh_e$ values. Additionally, $Re$ values do not provide a clear boundary between the two regimes. This behavior highlights the complexity of the problem and the importance of selecting an appropriate combination of parameters to distinguish the different impact regimes. Each simulation required between 6 and 10 hours to complete on an Intel(R) Xeon(R) CPU E5-2690 \@ 2.90GHz.

Since numerical simulations are often computationally expensive, especially at high spatial and temporal resolutions, we explored deep learning models as a promising alternative for accelerating the prediction of the temporal evolution of the fluid interface.

\subsection{Architectures}

The main objective is to define an architecture capable of reducing the computational costs of full numerical simulations while preserving sufficient physical fidelity to distinguish between spreading and bouncing regimes. We tested a simpler architecture, such as the Multilayer Perceptron (MLP), and the Video Vision Transformer (ViViT), a more complex architecture that enables the learning of global spatio-temporal dependencies.

\subsubsection{Multilayer Perceptron}

In the Multilayer Perceptron (MLP), signals are transmitted within a network of neurons in a single direction flowing from the input layer, which receives the input data, through one or more hidden layers, which process the data, to the output layer, which generates an approximation of the desired output.

In summary, the MLP is a function approximator capable of learning complex mappings between inputs and outputs through linear and nonlinear transformations. Considering an input vector $\mathbf{x} = [x_1, x_2, \dots, x_n]$ representing a flattened image, the transformation in the first hidden layer can be expressed as:

\begin{equation}
    \label{eq:1_mlp}
    \mathbf{h_1} = f(\mathbf{W_1} \mathbf{x} + \mathbf{b_1}),
\end{equation}
where $\mathbf{W}_1$ is the weight matrix connecting the input layer to the first hidden layer, $\mathbf{b}_1$ is the bias vector for the first hidden layer and $f$ is an activation function (for example, sigmoid, tanh, or ReLU), which transforms the resulting values. Equation \eqref{eq:1_mlp} is then repeated for each hidden layer $\mathbf{h_n}$ by substituting $x$ with the output from the previous hidden layer $\mathbf{h_{n-1}}$. After one or more hidden layers, the output layer is computed by:

\begin{equation}
    \tilde{\mathbf{y}} = g(\mathbf{W}_{n+1} \mathbf{h_n} + \mathbf{b}_{n+1}),
\end{equation}
in which $g$ is the activation function used in the output layer (sigmoid in our regression problem). The learning process in an MLP involves adjusting the weight matrices and bias vectors in order to minimize a loss function, which measures the discrepancy between the predicted output $\tilde{\mathbf{y}}$ and the expected value $\mathbf{y}$. In this work, optimization was performed via \textbf{backpropagation} combined with the Adaptive Moment Estimation (Adam) optimization algorithm.

\begin{figure}[htpb]
    \centering
    \includegraphics[width=0.9\linewidth]{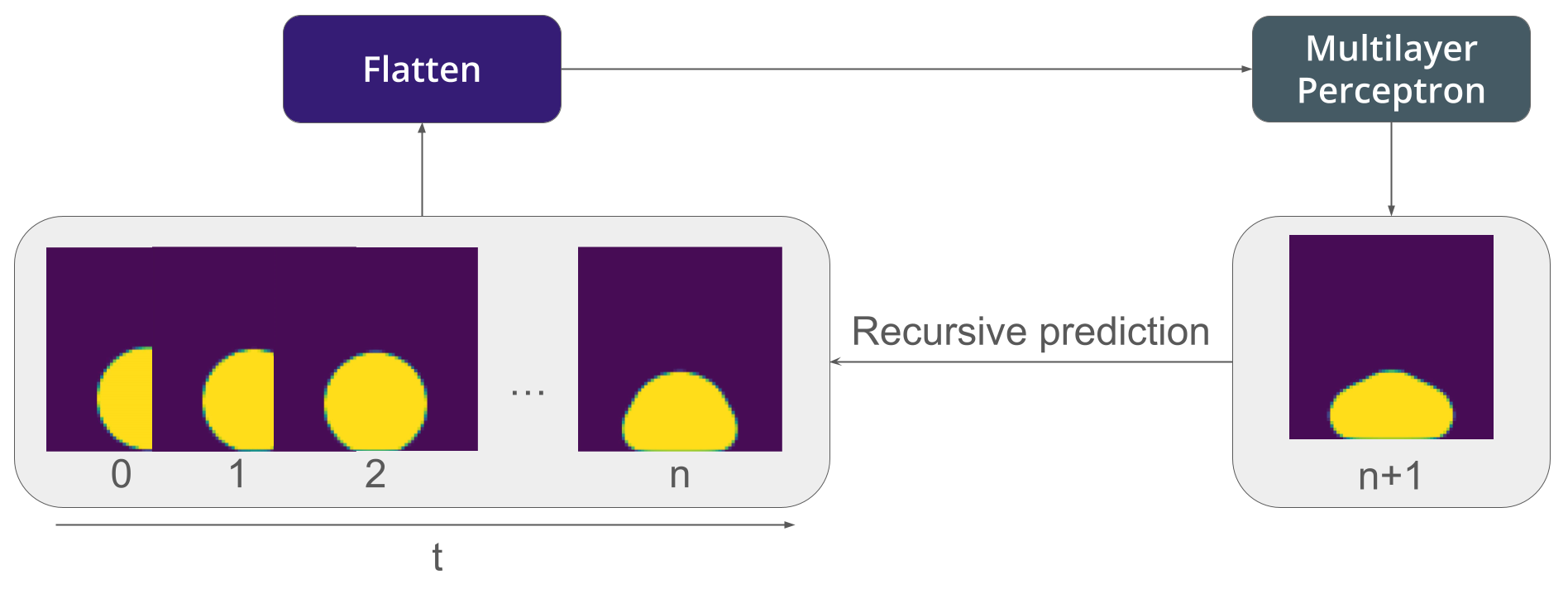}
    \caption{General architecture of the implemented MLP.}
    \label{fig:1_mlp}
\end{figure}

An initial context is used to predict a future time step and then, recursively, the model adds the new predicted data to predict further in time. The predicted output $\tilde{\mathbf{y}}$ is the flattened approximation of the volume fractions. Fig. \ref{fig:1_mlp} shows a simplified representation of the architecture for the prediction of a single time step, although multiple steps can be predicted simultaneously. The loss function then measures the distance between the values generated as output and the desired outcomes, guiding the optimizer to adjust the weights and minimize the prediction error.

\subsubsection{Video Vision Transformer}

The Vision Transformer architecture, introduced in \cite{Alexey2021}, is an adaptation of the transformer architecture \cite{Vaswani2017} for digital image processing. It largely relies on the mechanisms of the original Transformer, with modifications to adapt the image input so that it can be treated as sequential data.

To process and extract information from the data, the transformer mathematically mimics the psychological concept of self-attention. Using the self-attention mechanism, the model is able to capture relationships between all elements in a sequence, regardless of their distance, and filter only the most relevant information.

\begin{figure}[htpb]
    \centering
    \includegraphics[width=0.6\linewidth]{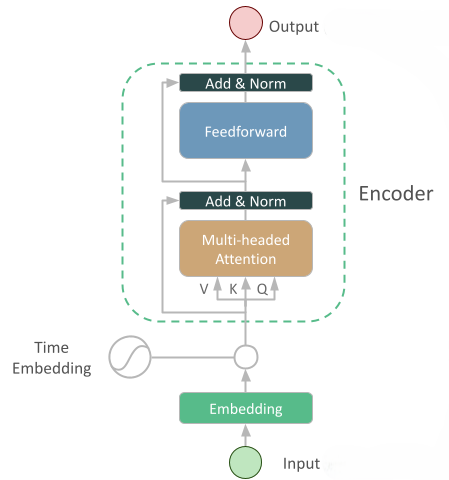}
    \caption{Basic transformer encoder architecture.}
    \label{fig:1_transformer}
\end{figure}

Its key components include: input embeddings, responsible for converting each input element into a continuous vector representation, time embedding, added to the input embeddings to retain sequence order information, multi-head attention, responsible for the self-attention mechanism, and feedforward layers for feature transformation. Specifically for our problem, the architecture is based solely on the encoder part of the transformer, shown in Fig. \ref{fig:1_transformer}.

\begin{description}
    \item[Input embedding:] to generate vector representations that capture relevant information for the attention layers, the input data first passes through dense layers. Since the generated output does not carry any positional information, they are then passed through a time embedding process.
    
    \item[Time embedding:] incorporates information about the position of each element in the sequence, helping the model ``understand'' the temporal structure of the data. This encoding is added to the embedding vectors before they are passed to the attention layers.
    
    The original work proposes a deterministic encoding based on sine and cosine functions of different frequencies. For each position $\text{pos}$ (which represents the index of the element in the sequence) and each dimension $i$ of the embedding vector, the value of the time embedding is defined by the following equations:

    \begin{align}
        & PE_{(pos, 2i)} = \sin \left( \frac{pos}{10000^{\frac{2i}{d}}} \right), \\
        & PE_{(pos, 2i+1)} = \cos \left( \frac{pos}{10000^{\frac{2i}{d}}} \right),
    \end{align}
    where $d$ represents the size of the embedding vector. The even dimensions of the vector ($2i$) receive values based on the sine function, while the odd dimensions ($2i+1$) use the cosine function, so that each dimension of the encoding varies according to a different frequency.
    
     Consequently, for each position $pos$, a vector $PE_{pos}$ is obtained and added directly to the corresponding embedding vector, resulting in the effective input vector for the transformer.
    
    \item[Multi-head Attention (MHA):] the attention mechanism is responsible for allowing each element in a sequence to attend to the others according to their relevance within the overall context of the problem.

    The mathematical implementation of the self-attention mechanism is performed by the Multi-head Attention (MHA) layer. The main objective of the MHA is to generate an attention filter that extracts important features and relates each element of a sequence to the others. To achieve this, the layer receives as input three matrices produced by the initial dense layers: \textbf{Query (Q)}, which acts as a ``question'' about how each element should attend to the others, \textbf{Key (K)}, which serves as a descriptor of each sequence element, determining the degree of relevance of one element to each of the others, and \textbf{Value (V)}, which contains the information that will be filtered by the relationships established between Queries and Keys.

    The process involves calculating how well \textbf{K} matches \textbf{Q} (degree of similarity) and, based on that, creating an attention matrix to filter elements of \textbf{V}. The proximity between \textbf{Q} and \textbf{K} is calculated using cosine similarity adapted to the matrix case:
    
    \begin{equation}
        \label{eq:sim_cosseno}
        \text{sim}_{cos} (Q, K) = \frac{Q K^T}{\sqrt{d}}
    \end{equation}
    where $d_k$ is the number of rows in the matrix \textbf{K}. The result of this operation produces a similarity matrix among the elements of a sequence, also called the attention filter. When passed through a softmax layer, this filter generates a probability distribution with values ranging from 0 to 1, in which the sum of each row equals 1:
    
    \begin{equation}
        \label{eq:softmax}
        \text{softmax}(x_i) = \frac{e^{x_i}}{\sum_{j = 1}^{n} e^{x_j}} 
    \end{equation}

    This operation represents the degree of relevance between elements in a sequence. Finally, the attention filter is multiplied by \textbf{V} to extract the most relevant information from the input data. The equation representing the entire attention calculation process described above is given by:
    
    \begin{equation}
        \label{eq:atencao}
        \text{Attention}(Q, K, V) = \text{softmax} \left ( \frac{QK^T}{\sqrt{d_k}} \right )V
    \end{equation}

    This process can be repeated multiple times by using more than one head. Since the values for \textbf{Q}, \textbf{K}, and \textbf{V} are influenced by the initial dense layers, each head can focus on a different feature according to what is deemed relevant during training. In the end, the results from all heads are concatenated and passed through a linear layer to reduce dimensionality and compile all the information extracted by the MHA.

    \item[Feedforward layers:] captures other complex patterns and enriches the generated representations. Its architecture is composed of three layers, with a ReLU activation layer between two linear activation layers. The ReLU activation function is responsible for removing negative values and keeping positive ones unchanged. 

    \item[Residual connections:] mechanism that allows the input of a layer to be added directly to its output, forming a shortcut around one or more layers, reinforcing signals and preventing the vanishing gradient problem. Mathematically, it can be defined as:

    \begin{equation}
        y = F(x) + x
    \end{equation}
    where $F(x)$ is the transformation of the original input $x$ by one or more layers.

    \item[Layer normalization:] reduces the range of values outputted by the neurons, making training faster and more stable.

\end{description}

\begin{figure}[htpb]
    \centerline{
    \includegraphics[width=0.8\linewidth]{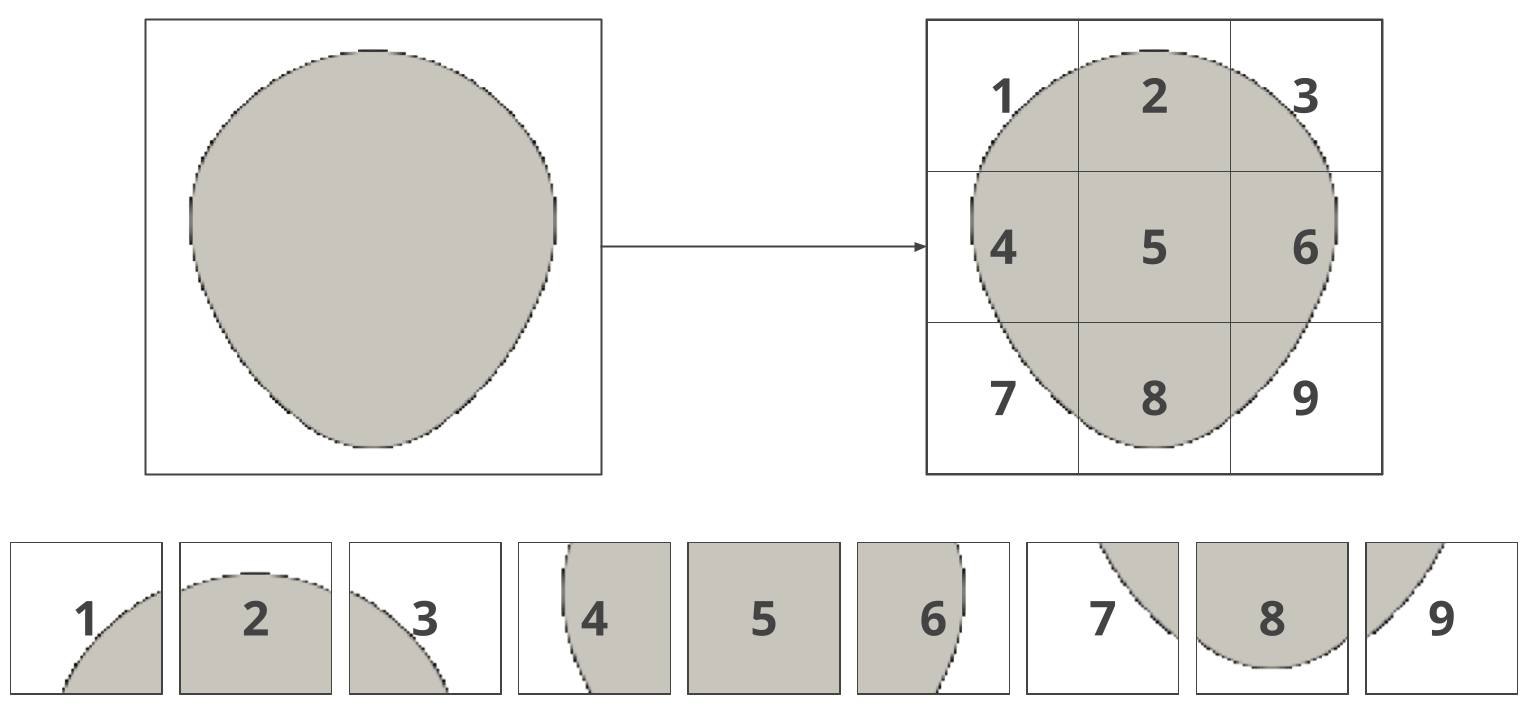}
    }
    \caption{The image is divided into non-overlapping patches, which are then rearranged into an ordered sequence from top to bottom and from left to right.}
    \label{fig:1_patch_embedding}
\end{figure}

To adapt image inputs for the input embedding, we first slice the images into blocks called patches. After this division, the blocks are indexed row by row, from left to right, and treated as a sequence of elements. The process is illustrated in Fig. \ref{fig:1_patch_embedding}.

\begin{figure}[htpb]
    \centering
    \begin{subfigure}{0.7\linewidth}
        \centering
        \includegraphics[width=1.0\linewidth]{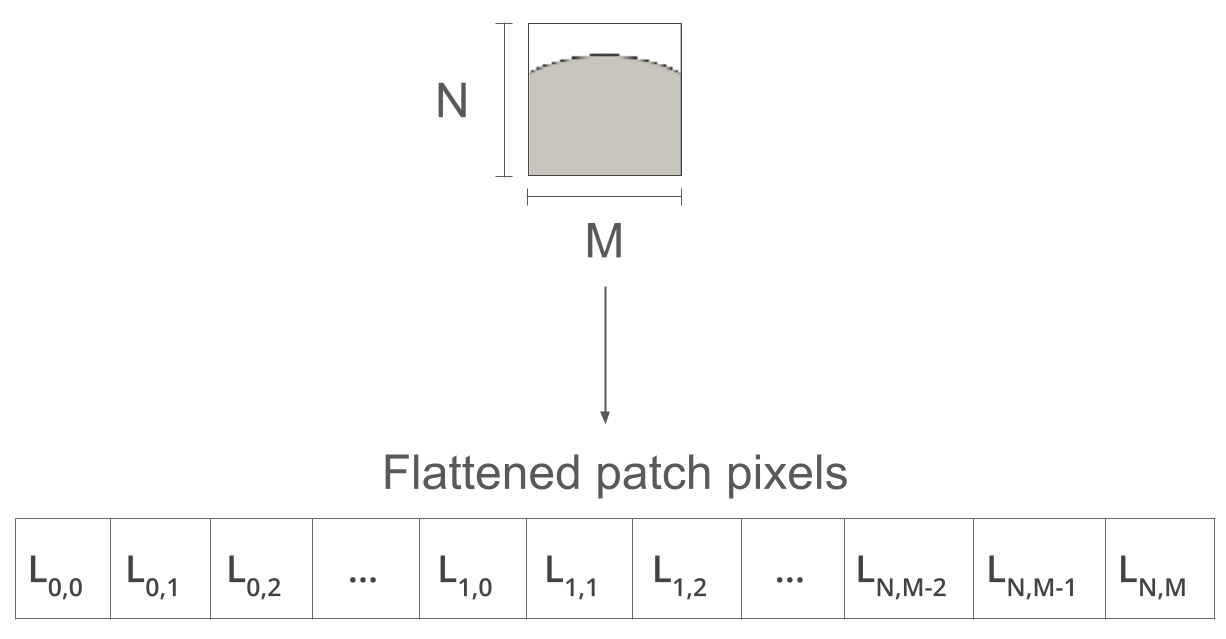}
        \caption{Flattening an image patch. Each pixel value $L_{i, j}$, where $i$ and $j$ denote, respectively, the row and column indices of the patch, is rearranged into a one-dimensional vector.}
        \label{fig:2_patch_embedding}
    \end{subfigure}
    \begin{subfigure}{0.7\linewidth}
        \centering
        \includegraphics[width=1.0\linewidth]{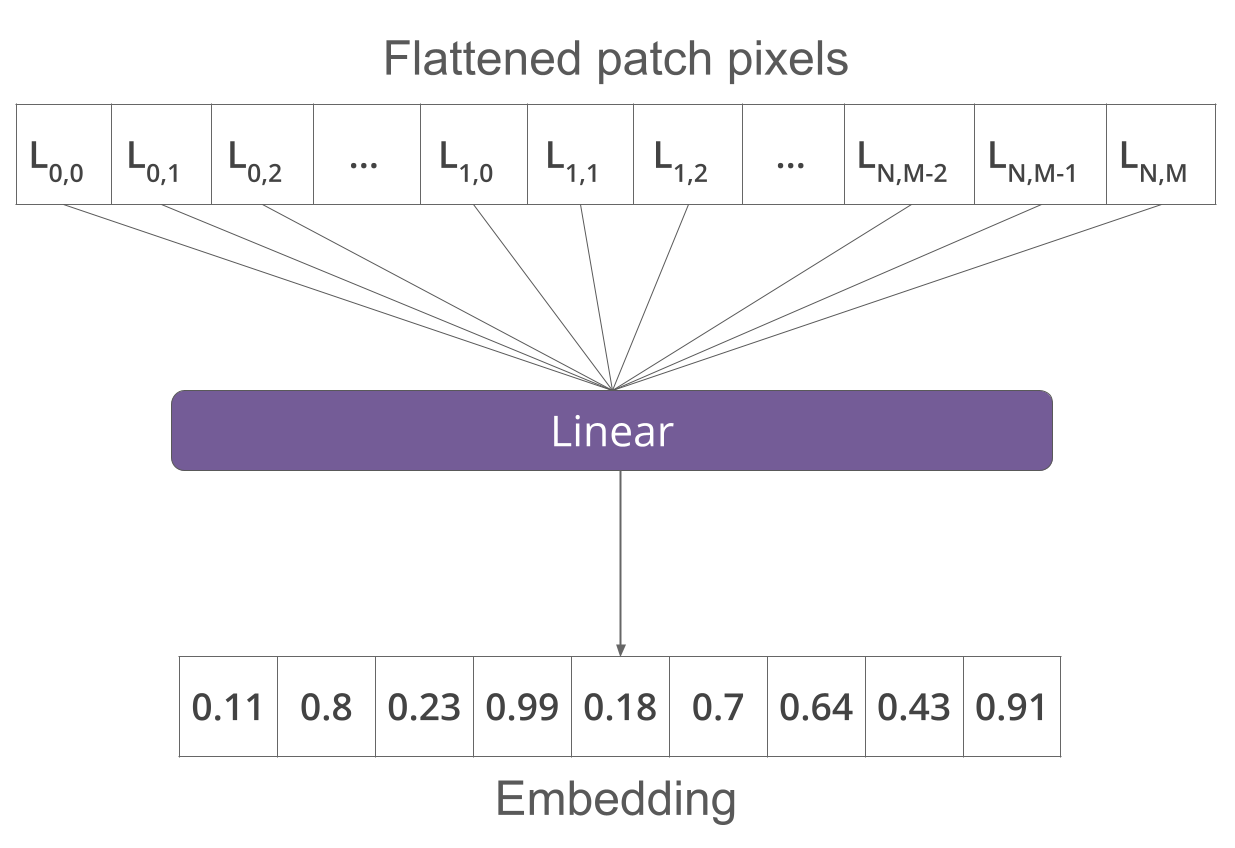}
        \caption{Example of embedding generation using a linear layer.}
        \label{fig:3_patch_embedding}
    \end{subfigure}
    \caption{Process of converting an image patch into an embedding vector.}
    \label{fig:patch_embedding}
\end{figure}

This image patching technique is used to define the spatial dependency relationships between different regions of an image. After this, the input embedding encodes them into lower-dimensional vectors. First, for each patch, the values of the pixels are flattened to a vector representation as shown in Fig. \ref{fig:2_patch_embedding}. Next, the generated vector passes through a linear layer, which processes the data and reduces its dimensionality, as shown in Fig. \ref{fig:3_patch_embedding}.

Since the order of the patches is important to establish the structural sequence of the image, a position-based encoding is added to each embedding. The attention mechanism then allows the architecture to extract global spatial patterns between all elements in a sequence, regardless of their distance, and to filter only the most relevant information to predict the next step. 

\begin{figure}[htpb]
    \centerline{
    \includegraphics[width=0.9\linewidth]{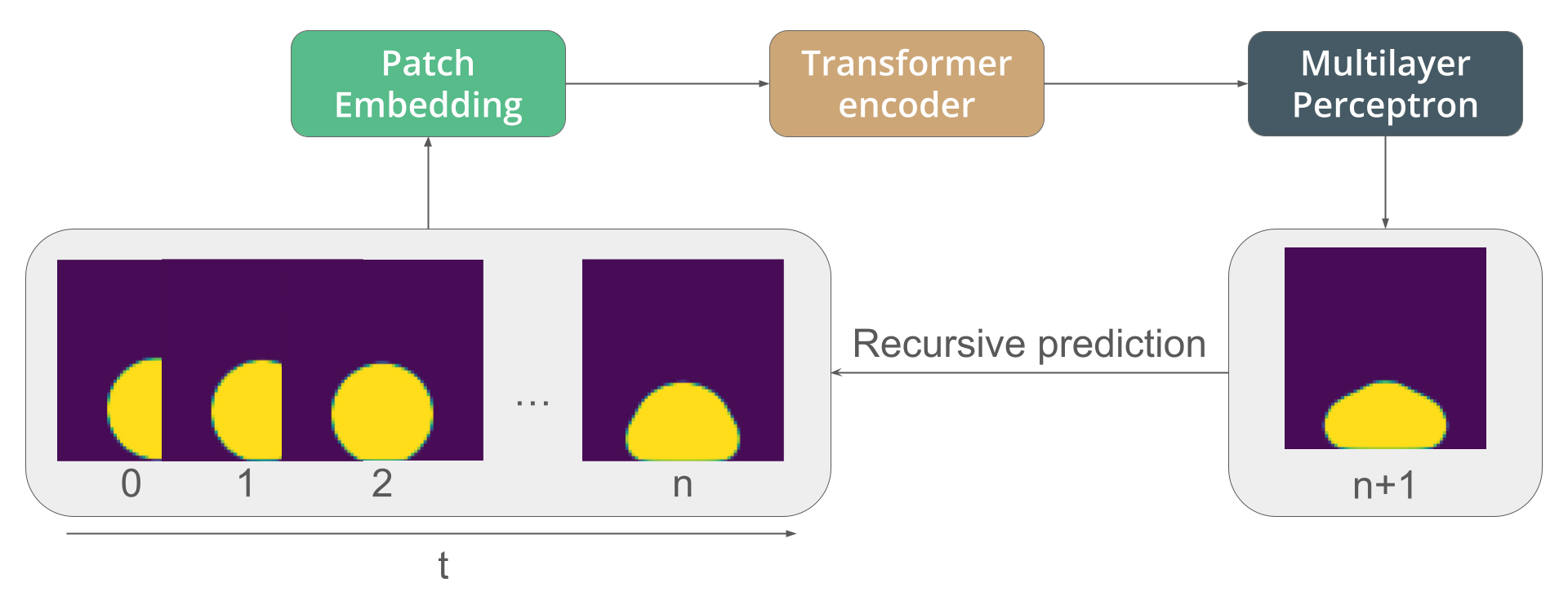}
    }
    \caption{General architecture of the implemented ViViT.}
    \label{fig:1_vivit}
\end{figure}

To adapt the ViT architecture to the task of predicting the next time steps using an initial context, we have implemented a Video Vision Transformer (ViViT) modification proposed by \cite{Arnab2021}. Instead of passing only one image, we pass a context of $n$ time steps and apply patch embedding to each of them sequentially. Next, the transformer encoder applies the attention mechanism to the embeddings generated by all patches from all images at the same time, granting not only global spatial relationships, but also temporal relationships. A representation of the general architecture adopted is shown in Fig. \ref{fig:1_vivit}, illustrating the prediction of a single time step, although multiple time steps can also be predicted at the same time.

\section{Results} \label{sec:results}

In this section, we describe how the dataset is divided into training, validation, and test sets, introduce the evaluation metrics used to assess model performance, and present and analyze the results obtained for each architecture.

\subsection{Training, validation, and test split}

To make the training process more challenging, we reserved some $We$ values exclusively for validation (to monitor overfitting) and testing (to evaluate the models on extrapolation tasks), since this is one of the parameters that most changes the bouncing behavior. A low Weber number indicates that surface tension dominates over inertial forces, causing the droplet to resist deformation and favoring bouncing upon impact, while a high Weber number indicates stronger inertial forces, usually leading to deformation and spreading over the solid surface.

The validation and test sets correspond to $We \in [0.5, 40]$, while the training set is defined as $We \in [0.5, 31.2222]$. This allows us to evaluate the ability of the trained models to both interpolate within the training range and extrapolate beyond it. In total, 126 parameter combinations are used for training (63 bouncing and 63 spreading), 27 for validation (20 bouncing and 7 spreading) and 27 for testing (7 bouncing and 20 spreading).

For each sample, three input $\rightarrow$ output configurations were evaluated: 50 time steps to predict the next step (50 $\rightarrow$ 1), 50 steps to predict the subsequent 50 steps (50 $\rightarrow$ 50), and 100 steps to predict the subsequent 100 steps (100 $\rightarrow$ 100).

\subsection{Evaluation metrics}

To evaluate the performance of our models, we adopted a set of standard evaluation metrics that assess four different aspects: predictive performance, measured by the $\text{R}^2$ score \eqref{eq:r2} and the Root Mean Square Error (RMSE) \eqref{eq:rmse}, the quality of the generated images, with the Structural Similarity Index Measure (SSIM) \eqref{eq:ssim}, the preservation of physical properties, evaluated through the Mean Absolute Error of the horizontal diameter ($\text{MAE}^{(\text{HD})}$), vertical diameter ($\text{MAE}^{(\text{VD})}$), y-axis center of mass ($\text{MAE}^{(\text{CM})}$) and contact time ($\text{MAE}^{(\text{CT})}$) \eqref{eq:mae}, and the computational cost, assessed through the training time \eqref{eq:time} on an NVIDIA GeForce RTX 3080 Ti GPU and the number of trainable parameters.

\begin{align}
    & \text{R}^2 = 1 - \frac{\sum_{n=1}^{N_{\text{test}}} \sum_{t=1}^{T_\text{timesteps}} 
    \left(Y^{(n,t)} - \tilde{Y}^{(n,t)}\right)^2}{\sum_{n=1}^{N_{\text{test}}} \sum_{t=1}^{T_{\text{timesteps}}} \left(Y^{(n,t)} - \mu_{Y} \right)^2}, \label{eq:r2} \\
    & \text{RMSE} = \sqrt{\frac{1}{\sum_{n=1}^{N_{\text{test}}} T_{\text{timesteps}}} 
    \sum_{n=1}^{N_{\text{test}}} \sum_{t=1}^{T_{\text{timesteps}}} 
    \left(Y^{(n,t)} - \tilde{Y}^{(n,t)}\right)^2}, \label{eq:rmse} \\
    & \text{SSIM} = \frac{1}{\sum_{n=1}^{N_{\text{test}}} T_{\text{timesteps}}}\sum_{n=1}^{N_{\text{test}}} \sum_{t=1}^{T_{\text{timesteps}}}\frac{(2\mu_Y^{(n,t)} \mu_{\tilde{Y}}^{(n,t)} + C_1)
(2C_{Y\tilde{Y}}^{(n,t)} + C_2)}{((\mu_Y^{(n,t)})^2 + (\mu_{\tilde{Y}}^{(n,t)})^2 + C_1)
((s_Y^{(n,t)})^2 + (s_{\tilde{Y}}^{(n,t)})^2 + C_2)}, \label{eq:ssim} \\
    & \text{MAE}^{(\phi)} = \frac{1}{N_{\text{test}}} \sum_{n=1}^{N_{\text{test}}} \left| \phi^{(n)} - \tilde{\phi}^{(n)} \right|, \label{eq:mae} \\
    & \text{training time} = \sum_{i=1}^{N_{\text{epochs}}} t_{\text{epoch}}^{(i)}. \label{eq:time}
\end{align}
where $N_{\text{test}}$ is the number of samples, $T_{\text{time steps}}$ is the number of time steps in each sample, $Y^{(n, t)}$ and $\tilde{Y}^{(n, t)}$ denote, respectively, the expected and the predicted volume fractions for sample $n$ at time $t$, $\mu$ is the mean, $s^2$ is the variance, $C_{Y\tilde{Y}}$ is the covariance between expected and predicted images, $C_1$ and $C_2$ are constants to stabilize the division, $\phi$ represents one of the geometric properties (horizontal and vertical diameters, center of mass or contact time), $N_\text{epochs}$ is the total number of training epochs and $t_{\text{epoch}}^{(i)}$ is the training time for epoch $i$. 

The $R^2$ score measures, with a maximum value of 1, how well the neural network can predict unseen data by comparing its performance to a baseline model that simply predicts the mean of the data. The RMSE evaluates the pointwise difference in the VOF field, penalizing larger errors more strongly. The SSIM compares how similar two images are considering their brightness, contrast and structure, with a value of 1 being a perfect match. The MAE of the geometric properties captures the deviation of the predicted morphology from the expected morphology. All metrics are computed using the architectures whose hyperparameters were optimized through 10 iterations of the automatic hyperparameter optimization software framework Optuna \cite{Takuya2019}.

\subsection{Proposed architectures}

The metrics for each recursive prediction strategy and both explored architectures are shown in Table \ref{tab:metricas}. To better visualize the prediction behavior in each regime, we selected four test samples, two from each regime, and analyzed their geometric properties and field predictions at selected time steps. For the spreading regime, we selected one case of extrapolation in terms of the Weber number.

\begin{table}[htpb]
\centering
\caption{Evaluation metrics for predictive performance, image quality, preservation of physical properties, and computational cost for each approach. Green cells indicate the best performance for a given metric, while red cells indicate the worst performance.}
\label{tab:metricas}
\begin{tabular}{|c|cc|cc|cc|}
\hline
\multirow{3}{*}{\textbf{Metrics}} & \multicolumn{6}{c}{\textbf{Architectures}} \\ \cline{2-7}
 & \textbf{MLP} & \textbf{ViViT} & \textbf{MLP} & \textbf{ViViT} & \textbf{MLP} & \textbf{ViViT} \\ 
 & \textbf{50 $\rightarrow$ 1} & \textbf{50 $\rightarrow$ 1} & \textbf{50 $\rightarrow$ 50} & \textbf{50 $\rightarrow$ 50} & \textbf{100 $\rightarrow$ 100} & \textbf{100 $\rightarrow$ 100} \\
\hline
$\mathbf{R^2}$ \textbf{Score} & 0.8692 & 0.9025 & \cellcolor{red!25} 0.0407 & 0.9234 & 0.2876 & \cellcolor{green!25} 0.9249 \\
\textbf{RMSE} & 0.1184 & 0.1022 & \cellcolor{red!25} 0.3207 & 0.0906 & 0.2724 & \cellcolor{green!25} 0.0884 \\
\hline
\textbf{SSIM} & 0.9455 & \cellcolor{green!25} 0.9576 & 0.7653 & 0.9569 & \cellcolor{red!25} 0.0393 & 0.9183 \\
\hline
$\mathbf{MAE^{(HD)}}$ & 0.1728 & 0.1719 & 1.2211 & 0.1291 & \cellcolor{red!25} 1.4285 & \cellcolor{green!25} 0.1057 \\
$\mathbf{MAE^{(VD)}}$ & 0.1459 & 0.1072 & \cellcolor{red!25} 0.9742 & 0.1011 & 0.8898 & \cellcolor{green!25} 0.0923 \\
$\mathbf{MAE^{(CM)}}$ & 0.0823 & 0.0556 & 0.7743 & 0.0448 & \cellcolor{red!25} 0.8252 & \cellcolor{green!25} 0.0439 \\  
$\mathbf{MAE^{(CT)}}$ & 0.1185 & \cellcolor{green!25} 0.0037 & 1.3111 & 0.0067 & \cellcolor{red!25} 1.3215 & 0.0126 \\
\hline
\textbf{Training time (h)} & 6.9444 & \cellcolor{red!25} 12.0555 & \cellcolor{green!25} 2.0533 & 7.1333 & 2.8088 & 6.7589 \\
\textbf{Parameters (M)} & \cellcolor{green!25} 19.7461 & 40.3447 & 62.4780 & 28.0819 & 49.1597 & \cellcolor{red!25} 81.6619 \\
\hline
\end{tabular}

\vspace{0.2cm}

\footnotesize
MAE denotes the Mean Absolute Error. The superscripts indicate the evaluated geometric property: horizontal diameter (HD), vertical diameter (VD), y-axis center of mass (CM), and contact time (CT).
\normalsize

\end{table}

\begin{figure}[htpb]
    \centering
    \begin{subfigure}{1.0\linewidth}
        \centering
        \includegraphics[width=1.0\linewidth]{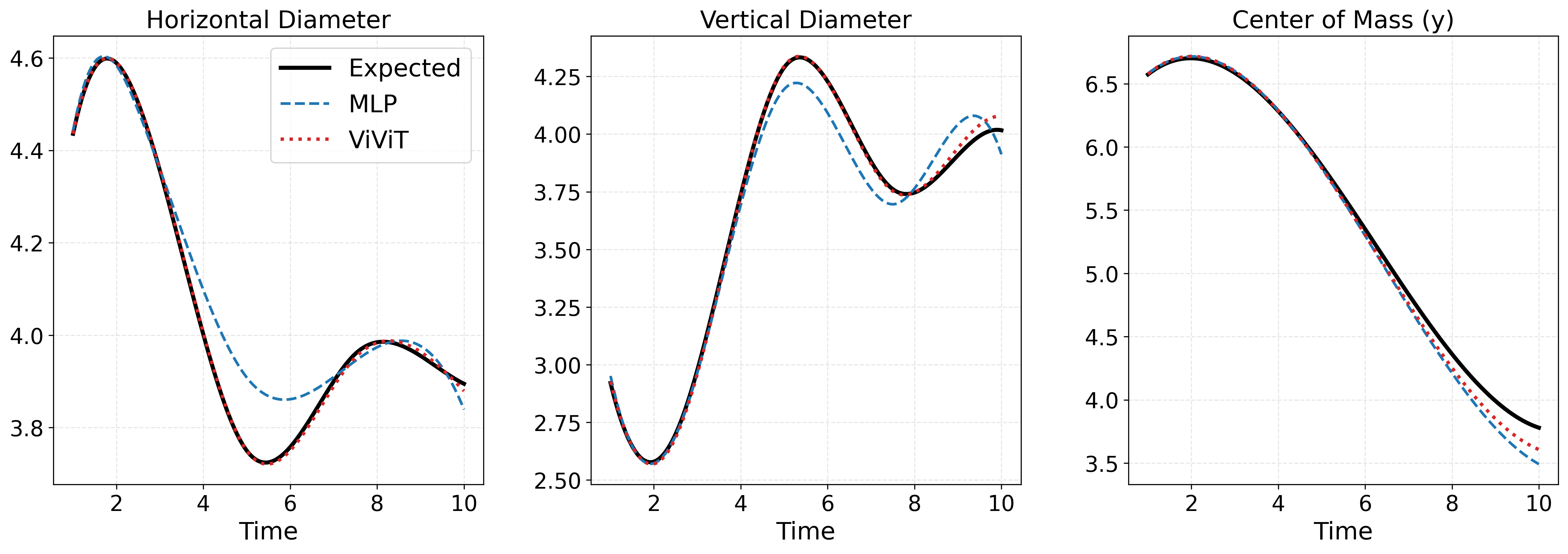}
        \caption{Bouncing sample with $Re = 2.5$, $We = 0.5$, $\beta = 0.9$ and $Wi = 3.0$.}
        \label{fig:metricas_bouncing_50_1}
    \end{subfigure}
    \begin{subfigure}{1.0\linewidth}
        \centering
        \includegraphics[width=1.0\linewidth]{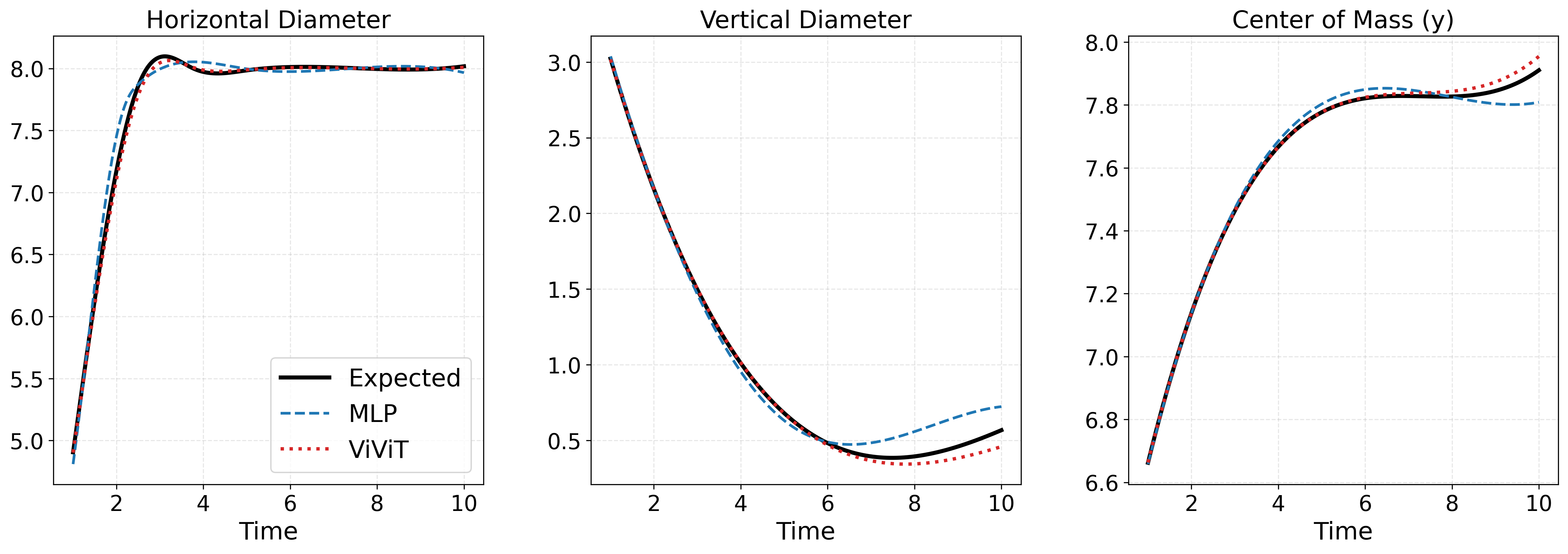}
        \caption{Spreading sample with $Re = 35.28$, $We = 40.0$, $\beta = 0.5$ and $Wi = 1.0$.}
        \label{fig:metricas_spreading_50_1}
    \end{subfigure}
    \caption{Comparison between expected and predicted values using MLP and ViViT (50 → 1). Predictions start at $t = 1$, as the initial 50 time steps are used as input.}
    \label{fig:metricas_50_1}
\end{figure}

\begin{figure}[htpb]
    \centering
    \begin{subfigure}{1.0\linewidth}
        \centering
        \includegraphics[width=1.0\linewidth]{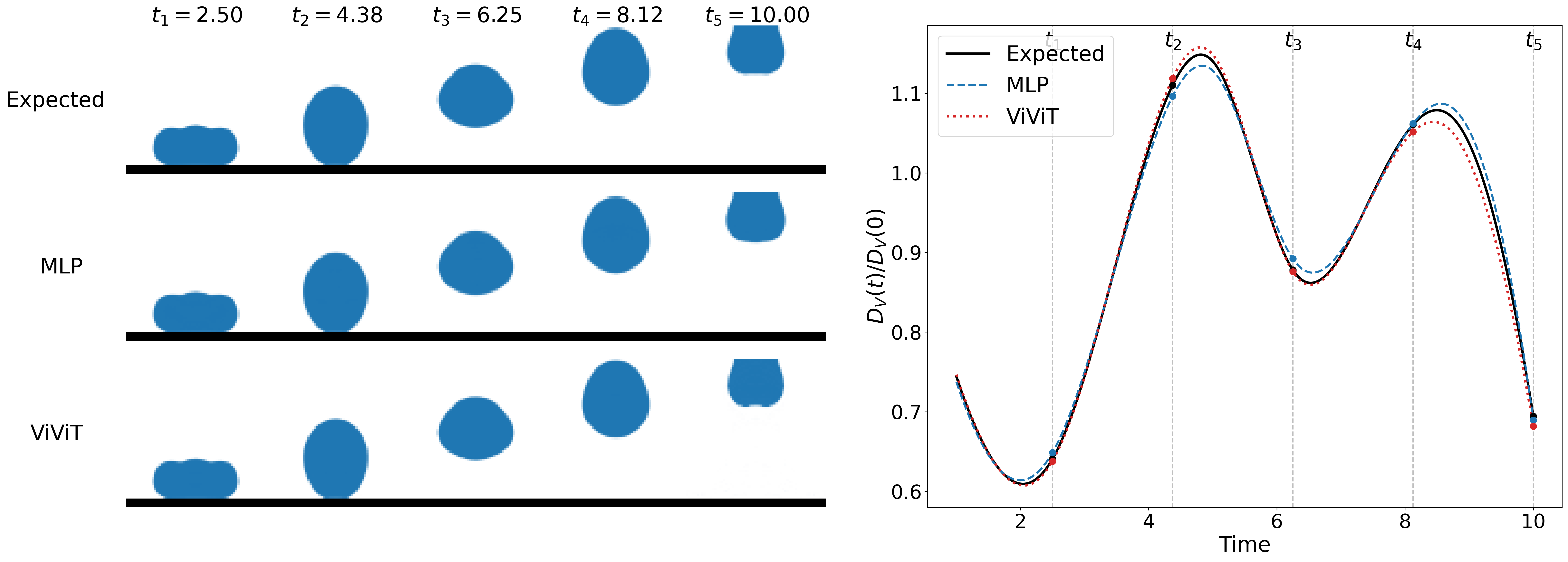}
        \caption{Bouncing sample with $Re = 68.06$, $We = 0.5$, $\beta = 0.1$ and $Wi = 3.0$.}
        \label{fig:combinacao_bouncing_50_1}
    \end{subfigure}
    \begin{subfigure}{1.0\linewidth}
        \centering
        \includegraphics[width=1.0\linewidth]{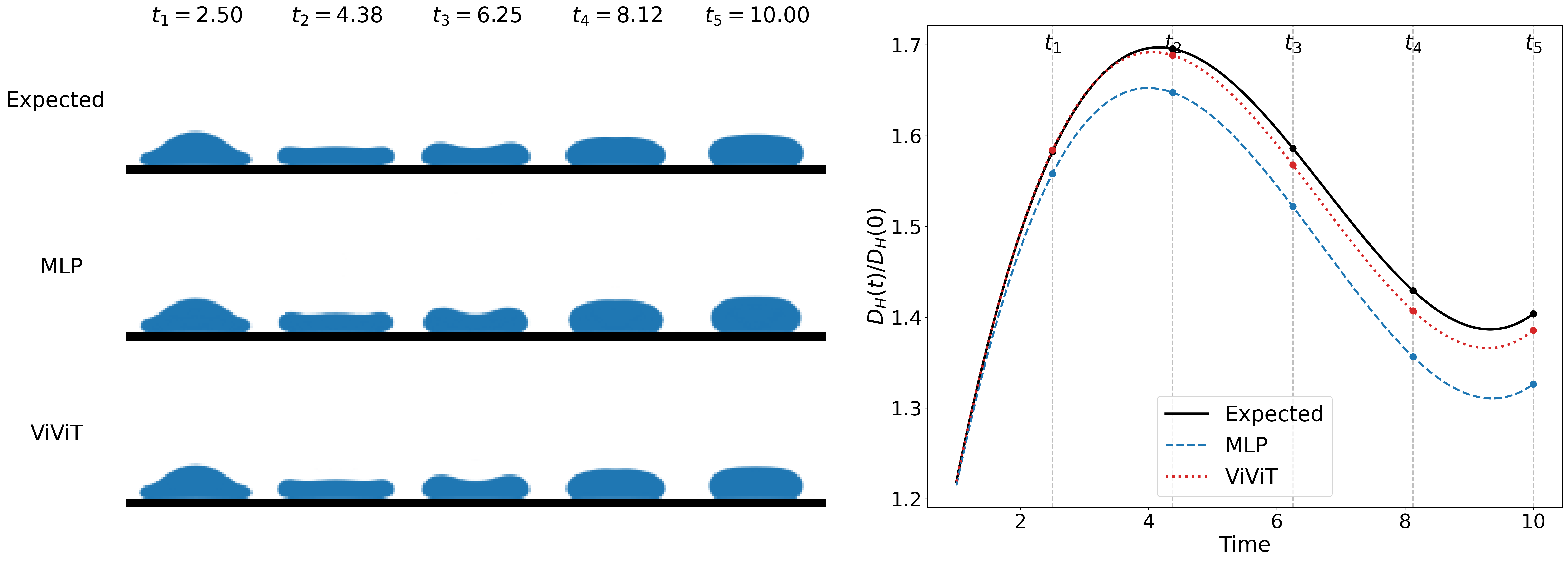}
        \caption{Spreading sample with $Re = 2.5$, $We = 40.0$, $\beta = 0.1$ and $Wi = 3.0$.}
        \label{fig:combinacao_spreading_50_1}
    \end{subfigure}
    \caption{Comparison between expected and predicted droplet dynamics using MLP and ViViT (50 → 1). Left column shows the volume fraction fields at five time steps ($t = 2.5, 4.38, 6.25, 8.12$ and $10.0$). Right column presents the corresponding normalized horizontal diameter $D_H(t)/D_H(0)$ for spreading cases or normalized vertical diameter $D_V(t)/D_V(0)$ for bouncing cases. Predictions start at $t = 1$, as the initial 50 time steps are used as input.}
    \label{fig:combinacao_50_1}
\end{figure}

For the 50 → 1 scenario, in which the prediction is performed step by step, both architectures still exhibit reasonable performances. In this context, the problem remains well behaved and error accumulation has not yet become a significant issue for the MLP. However, ViViT already consistently outperforms the MLP in terms of predictive performance, structural similarity and physical characteristics such as the contact time, for which it achieves the best results across all experiments. The main trade-off is the training time, which is nearly twice that of the MLP. This approach is computationally more expensive for both architectures because each new time step must be predicted individually, resulting in a larger training dataset. Fig \ref{fig:metricas_50_1} illustrates the evolution of the geometric properties for two selected samples, showing closer agreement between expected and predicted values for the ViViT model. Additional qualitative examples are presented in Fig. \ref{fig:combinacao_50_1} for a different set of parameters, where selected volume fraction fields are displayed together with the evolution of the normalized horizontal diameter for spreading cases and the normalized vertical diameter for bouncing cases.

\begin{figure}[htpb]
    \centering
    \begin{subfigure}{1.0\linewidth}
        \centering
        \includegraphics[width=1.0\linewidth]{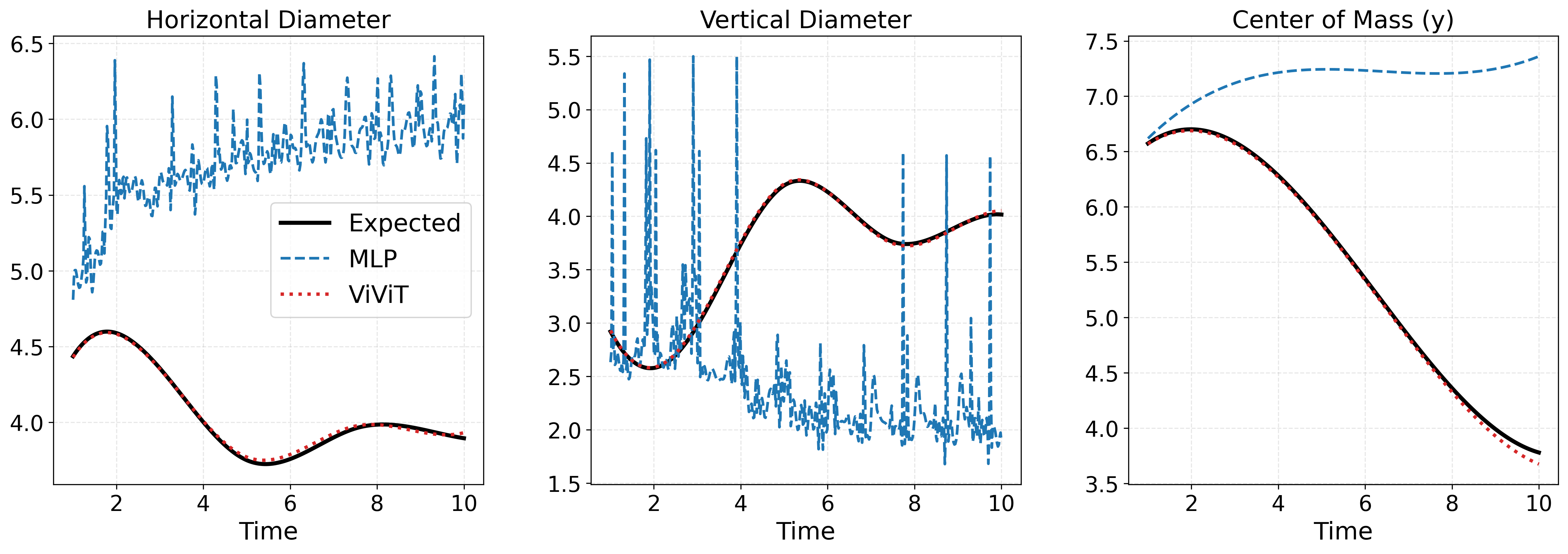}
        \caption{Bouncing sample with $Re = 2.5$, $We = 0.5$, $\beta = 0.9$ and $Wi = 3.0$.}
        \label{fig:metricas_bouncing_50_50}
    \end{subfigure}

    \begin{subfigure}{1.0\linewidth}
        \centering
        \includegraphics[width=1.0\linewidth]{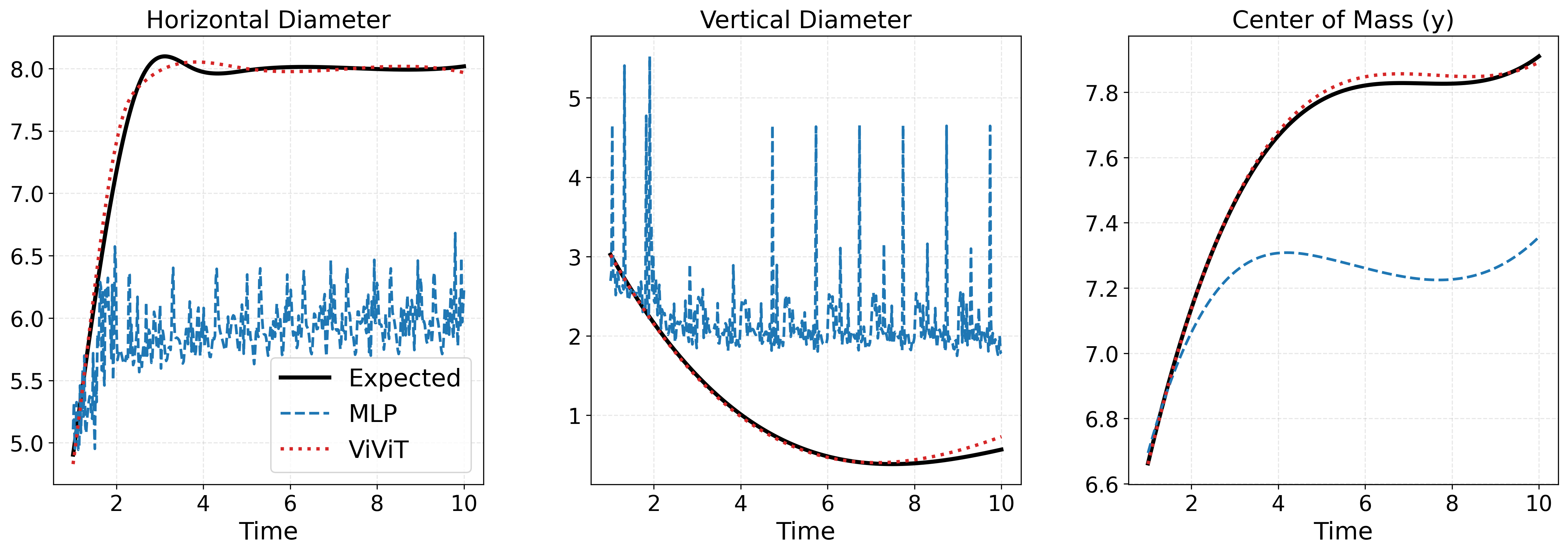}
        \caption{Spreading sample with $Re = 35.28$, $We = 40.0$, $\beta = 0.5$ and $Wi = 1.0$.}
        \label{fig:metricas_spreading_50_50}
    \end{subfigure}

    \caption{Comparison between expected and predicted values using MLP and ViViT (50 → 50). Predictions start at $t = 1$, as the initial 50 time steps are used as input.}
    \label{fig:metricas_50_50}
\end{figure}

\begin{figure}[htpb]
    \centering
    \begin{subfigure}{1.0\linewidth}
        \centering
        \includegraphics[width=1.0\linewidth]{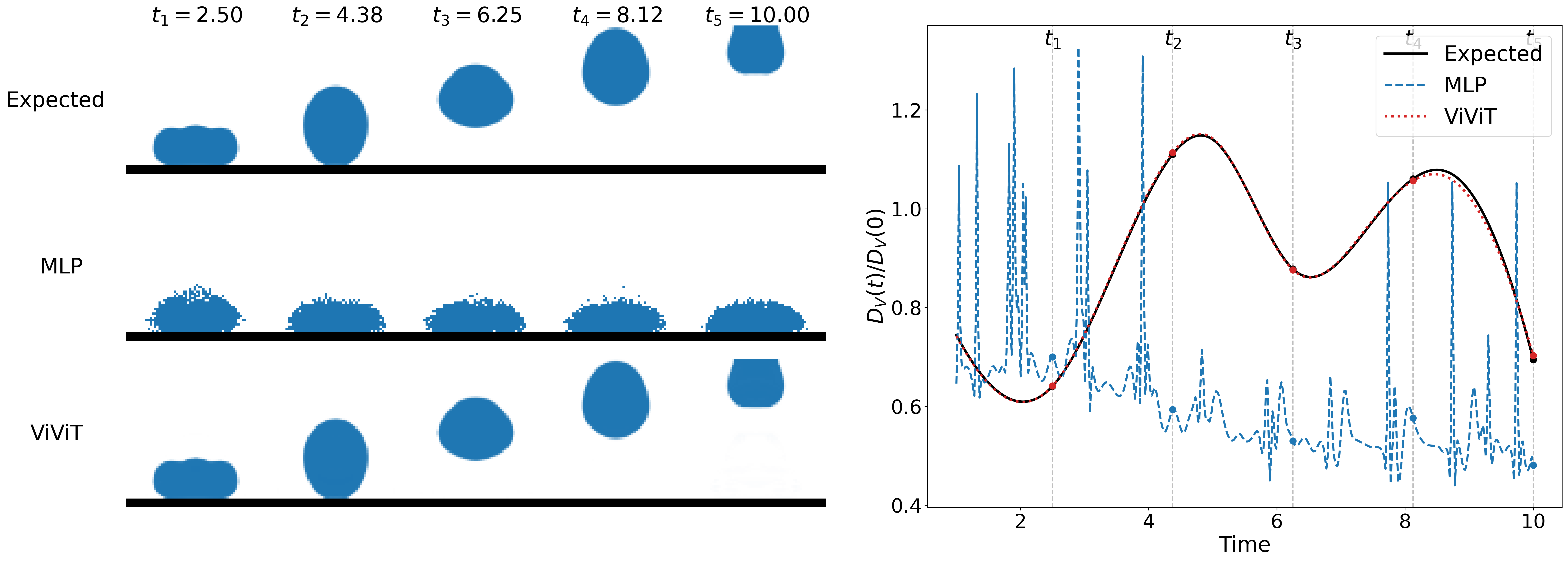}
        \caption{Bouncing sample with $Re = 68.06$, $We = 0.5$, $\beta = 0.1$ and $Wi = 3.0$.}
        \label{fig:combinacao_bouncing_50_50}
    \end{subfigure}
    \begin{subfigure}{1.0\linewidth}
        \centering
        \includegraphics[width=1.0\linewidth]{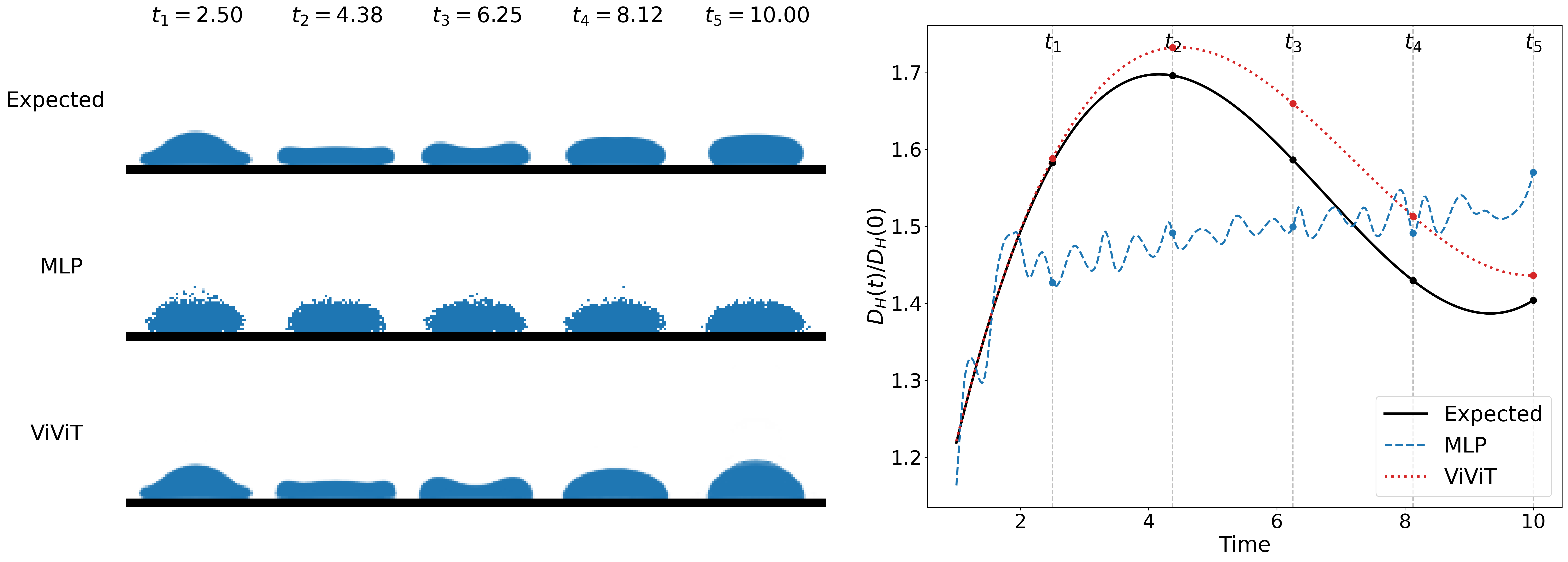}
        \caption{Spreading sample with $Re = 2.5$, $We = 40.0$, $\beta = 0.1$ and $Wi = 3.0$.}
        \label{fig:combinacao_spreading_50_50}
    \end{subfigure}
    \caption{Comparison between expected and predicted droplet dynamics using MLP and ViViT (50 → 50). Left column shows the volume fraction fields at five time steps ($t = 2.5, 4.38, 6.25, 8.12$ and $10.0$). Right column presents the corresponding normalized horizontal diameter $D_H(t)/D_H(0)$ for spreading cases or normalized vertical diameter $D_V(t)/D_V(0)$ for bouncing cases. Predictions start at $t = 1$, as the initial 50 time steps are used as input.}
    \label{fig:combinacao_50_50}
\end{figure}

By expanding the horizon of predictions, the differences between architectures become more noticeable. In the 50 → 50 case, the MLP collapses, with the $R^2$ score dropping close to zero and all geometric properties exploding in error, since almost all the predicted time steps captured by the model are just approximations of the mean behavior observed across all samples. The MLP is incapable of representing dependencies between future steps, treating the task as a direct regression without modeling the underlying dynamics. In contrast, the ViViT model not only maintains a stable performance, but also achieves better metrics for most characteristics, showing that both the values and the structure of the volume fields are well captured. This highlights the main advantages of using a transformer-based architecture, which is capturing long-range dependencies and, in the case of the ViViT modification, extending this to both temporal and spatial relationships. Fig \ref{fig:metricas_50_50} shows these results for two of the representative samples. Fig. \ref{fig:combinacao_50_50} further highlights the limitations of the MLP model, as both the predicted droplet morphology and the geometric evolution deviate significantly from the expected behavior.

\begin{figure}[htpb]
    \centering
    \begin{subfigure}{1.0\linewidth}
        \centering
        \includegraphics[width=1.0\linewidth]{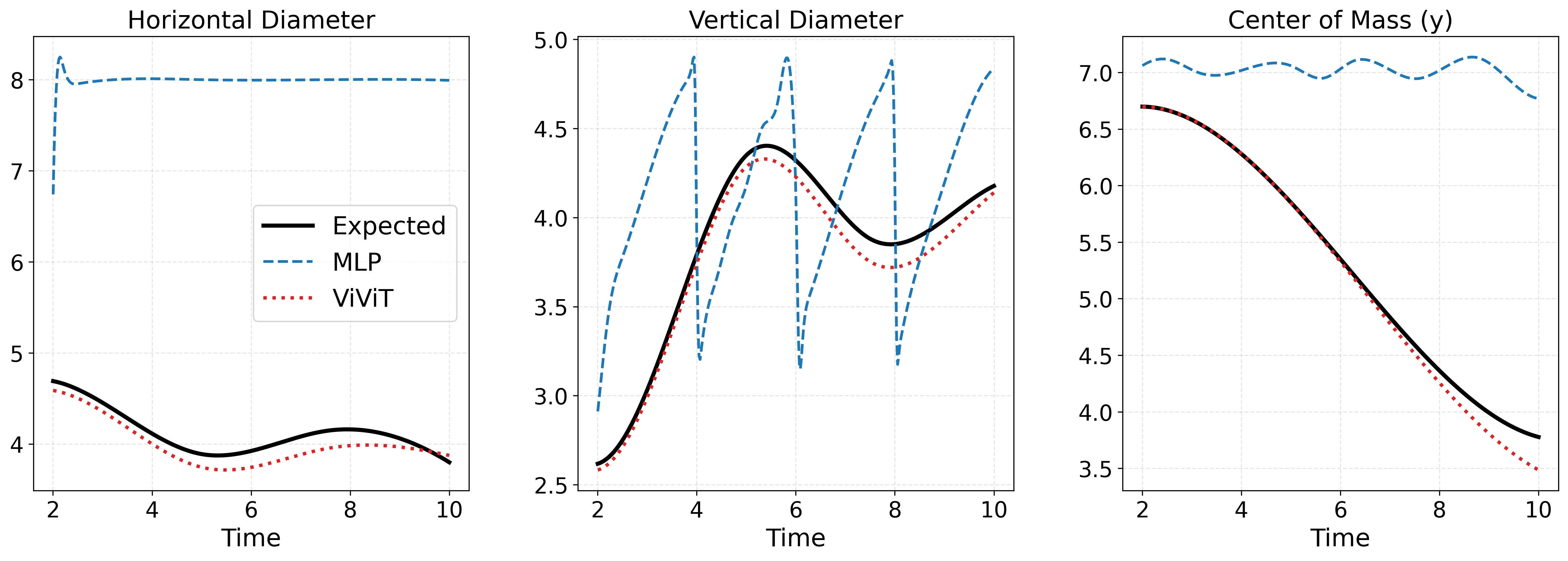}
        \caption{Bouncing sample with $Re = 2.5$, $We = 0.5$, $\beta = 0.9$ and $Wi = 3.0$.}
        \label{fig:metricas_bouncing_100_100}
    \end{subfigure}

    \begin{subfigure}{1.0\linewidth}
        \centering
        \includegraphics[width=1.0\linewidth]{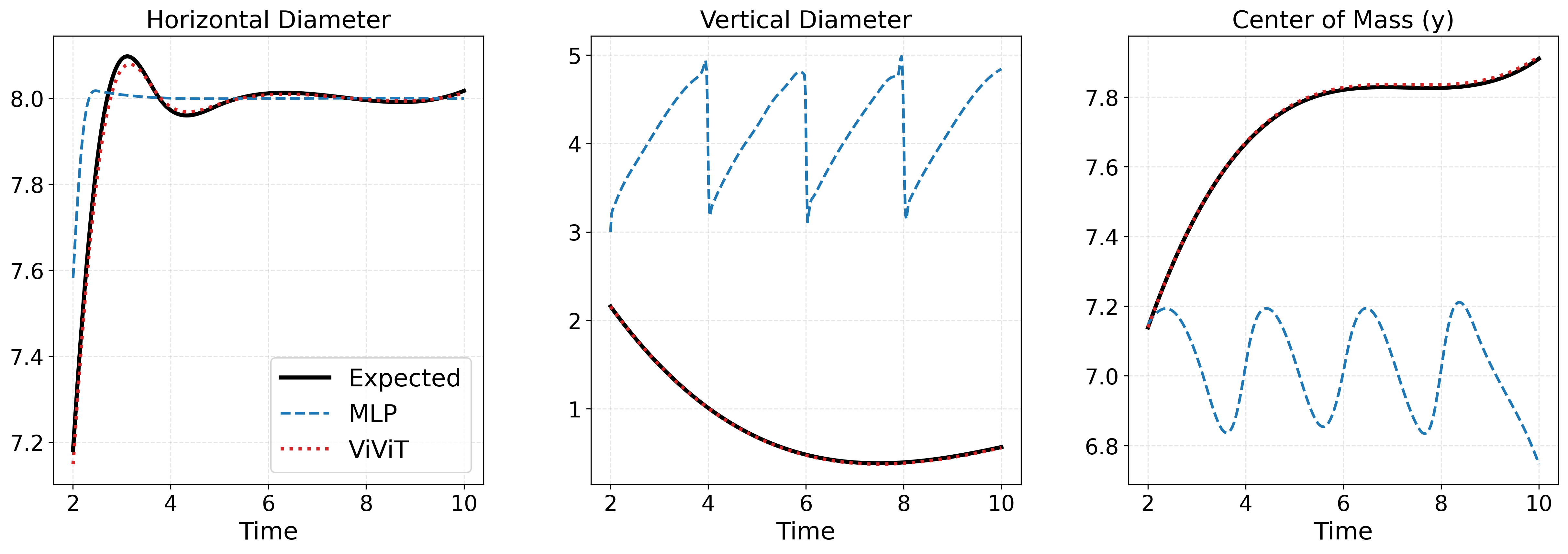}
        \caption{Spreading sample with $Re = 35.28$, $We = 40.0$, $\beta = 0.5$ and $Wi = 1.0$.}
        \label{fig:metricas_spreading_100_100}
    \end{subfigure}

    \caption{Comparison between expected and predicted values using MLP and ViViT (100 → 100). Predictions start at $t = 2$, as the initial 100 time steps are used as input.}
    \label{fig:metricas_100_100}
\end{figure}

\begin{figure}[htpb]
    \centering
    \begin{subfigure}{1.0\linewidth}
        \centering
        \includegraphics[width=1.0\linewidth]{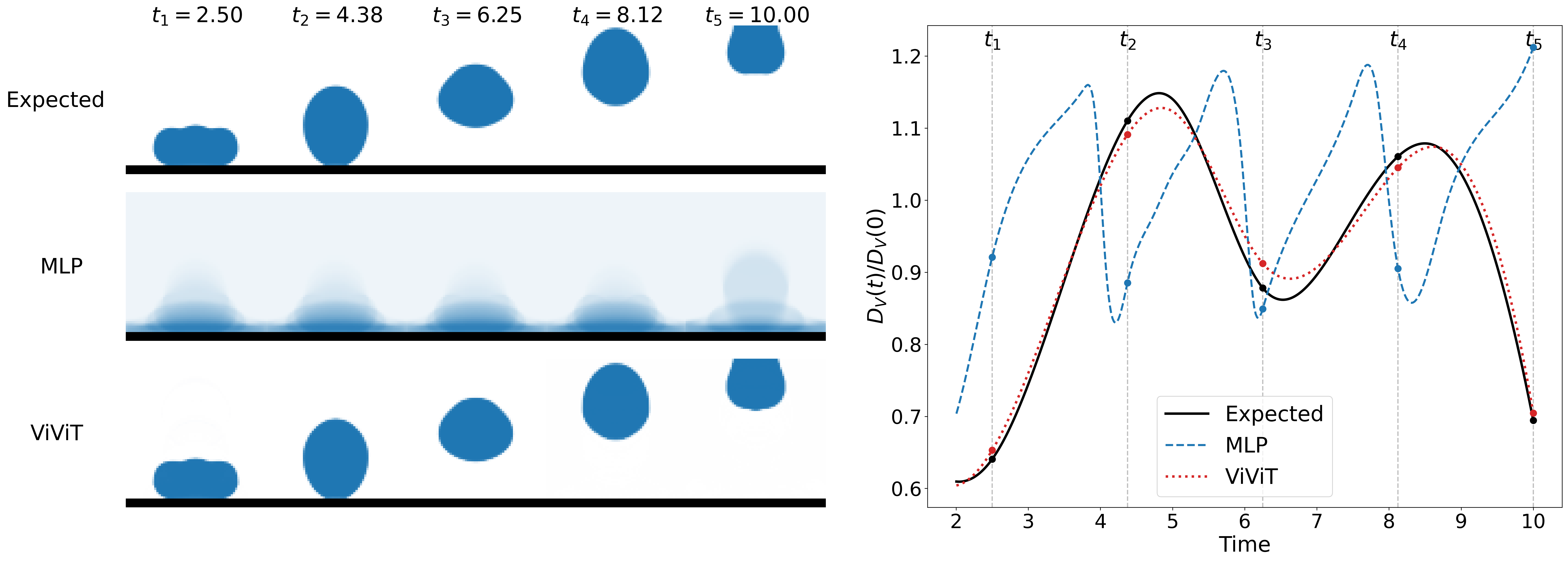}
        \caption{Bouncing sample with $Re = 68.06$, $We = 0.5$, $\beta = 0.1$ and $Wi = 3.0$.}
        \label{fig:combinacao_bouncing_100_100}
    \end{subfigure}
    \begin{subfigure}{1.0\linewidth}
        \centering
        \includegraphics[width=1.0\linewidth]{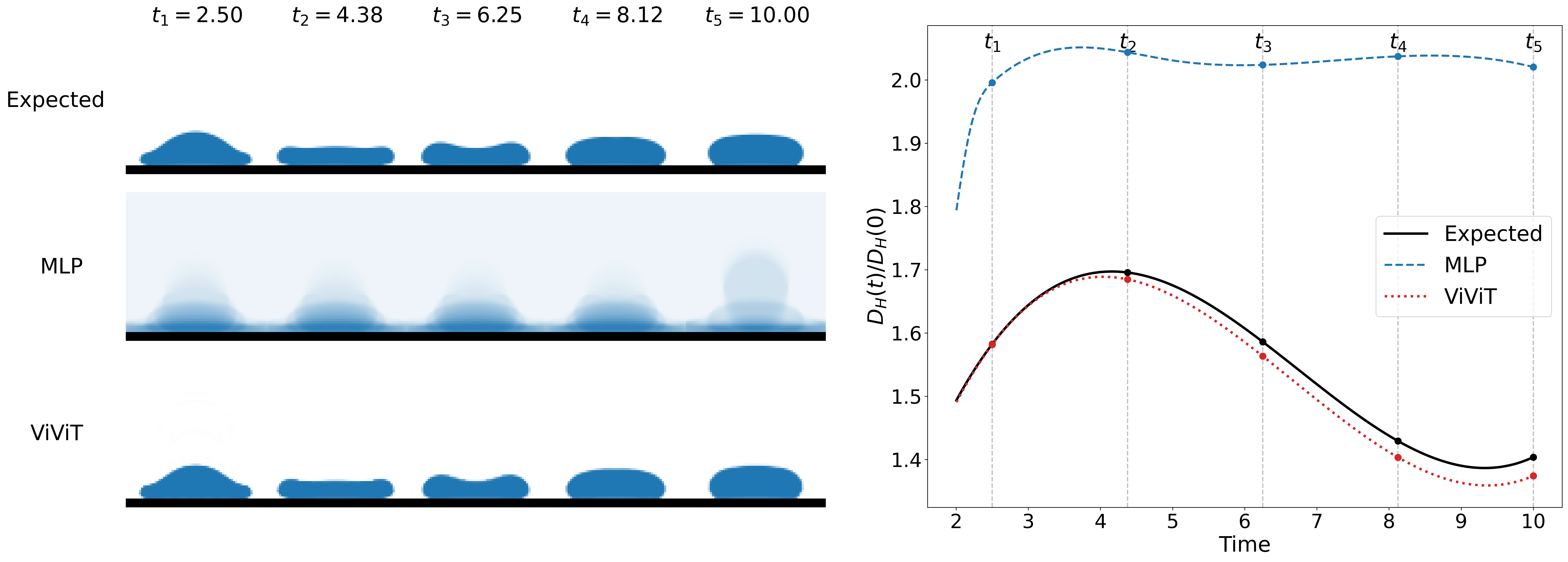}
        \caption{Spreading sample with $Re = 2.5$, $We = 40.0$, $\beta = 0.1$ and $Wi = 3.0$.}
        \label{fig:combinacao_spreading_100_100}
    \end{subfigure}
    \caption{Comparison between expected and predicted droplet dynamics using MLP and ViViT (100 → 100). Left column shows the volume fraction fields at five time steps ($t = 2.5, 4.38, 6.25, 8.12$ and $10.0$). Right column presents the corresponding normalized horizontal diameter $D_H(t)/D_H(0)$ for spreading cases or normalized vertical diameter $D_V(t)/D_V(0)$ for bouncing cases. Predictions start at $t = 2$, as the initial 100 time steps are used as input.}
    \label{fig:combinacao_100_100}
\end{figure}

Lastly, in the 100 → 100 scenario, the difference becomes even more pronounced. For the MLP, there are almost no structural similarities between the predicted and the expected volume fractions, and most geometric properties, such as the horizontal diameter, the center of mass and the contact time, exhibit the highest errors observed, indicating that the architecture is unable to maintain physical consistency. In comparison, the ViViT stabilizes even more, achieving lower prediction errors and better capturing the underlying dynamics, as observed through the geometric properties. Although Fig. \ref{fig:metricas_100_100} shows slightly worse predictions for the bouncing sample, the results for the spreading sample and the overall metrics indicate improved average performance. By analyzing Fig. \ref{fig:combinacao_100_100}, it becomes evident that the MLP predictions converge toward an average behavior learned from the training data, failing to preserve individual regime dynamics.

\begin{figure}[htpb]
    \centerline{
    \includegraphics[width=1.0\linewidth]{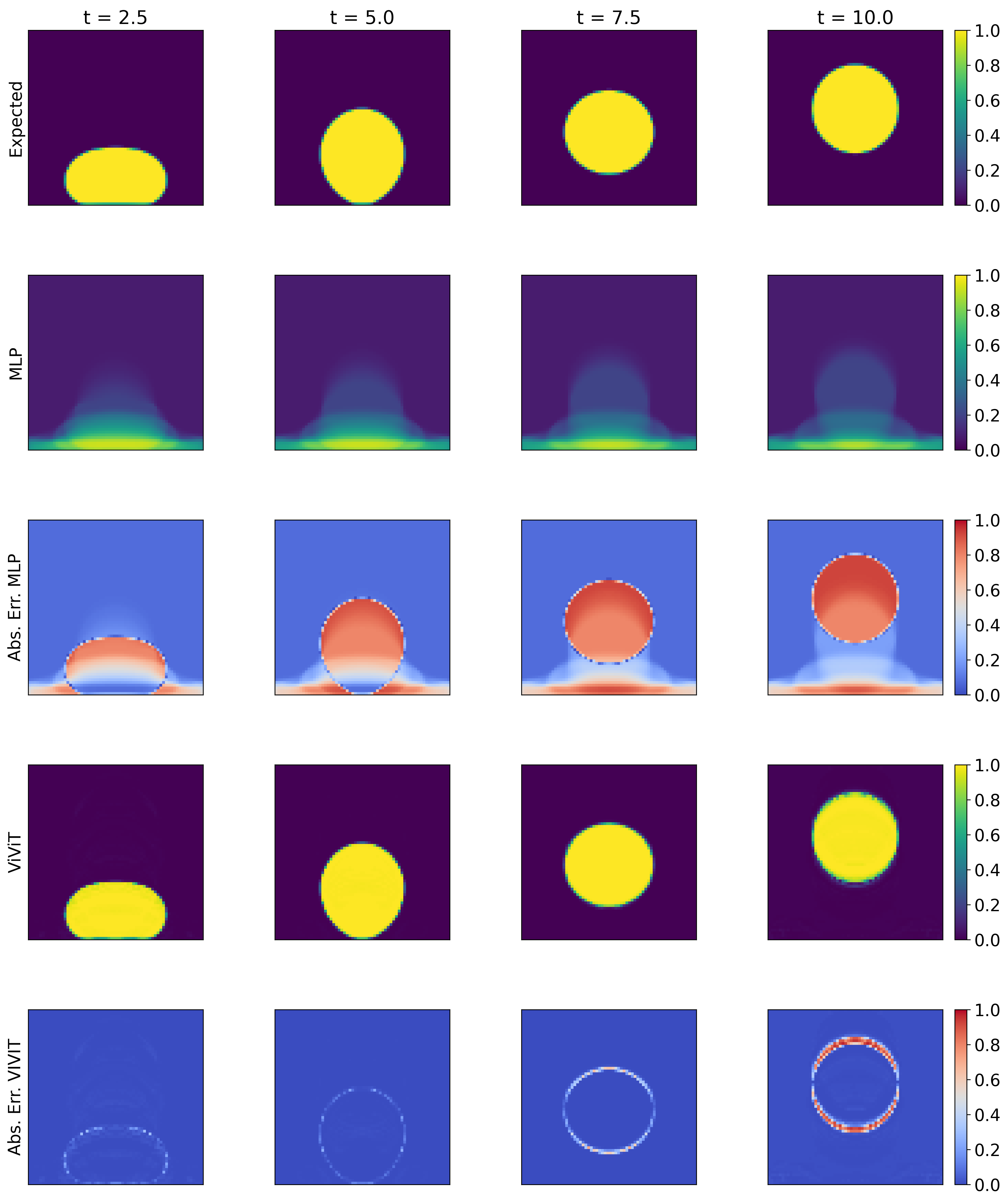}
    }
    \caption{Comparison of expected and predicted volume fraction fields obtained with MLP and ViViT at $t = 2.5$, $t = 5.0$, $t = 7.5$ and $t = 10.0$ for the bouncing sample with $Re = 2.5$, $We = 0.5$, $\beta = 0.9$ and $Wi = 3.0$.}
    \label{fig:snapshots_bouncing_100_100}
\end{figure}

\begin{figure}[htpb]
    \centerline{
    \includegraphics[width=1.0\linewidth]{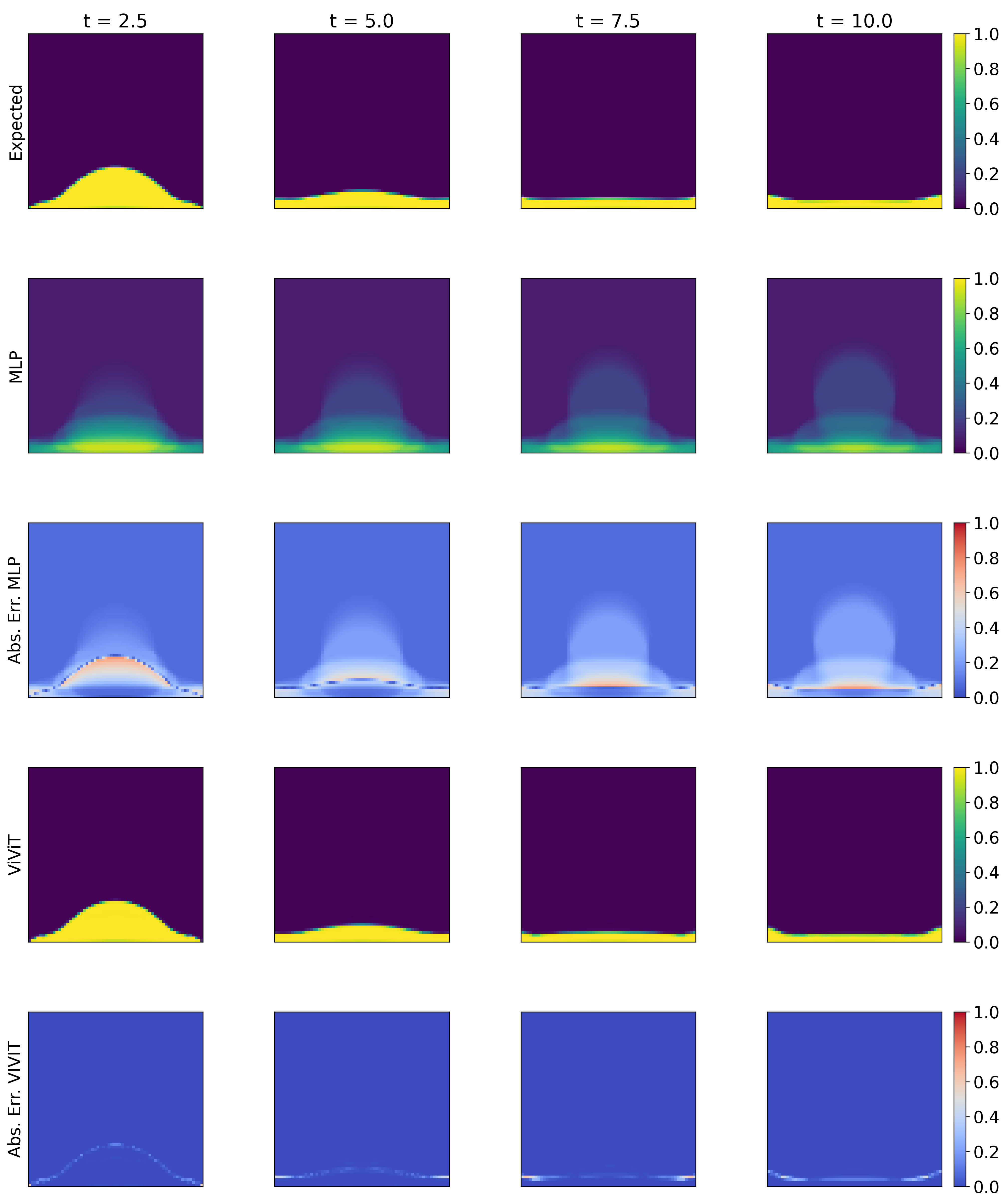}
    }
    \caption{Comparison of expected and predicted volume fraction fields obtained with MLP and ViViT at $t = 2.5$, $t = 5.0$, $t = 7.5$ and $t = 10.0$ for the spreading sample with $Re = 35.28$, $We = 40.0$, $\beta = 0.5$ and $Wi = 1.0$.}
    \label{fig:snapshots_spreading_100_100}
\end{figure}

Overall, although computationally more expensive, the ViViT-based experiments yielded the best results with greater robustness and scalability. In particular, the ViViT model that uses 100 time steps as input to predict the subsequent 100 time steps achieved the best performance and trade-off between accuracy and computational cost when compared to its MLP counterpart. Figs. \ref{fig:snapshots_bouncing_100_100} and \ref{fig:snapshots_spreading_100_100} present comparisons of the full volume fraction fields at selected time steps, demonstrating how the MLP fails to represent the differences between regimes, while the ViViT successfully captures them with lower error accumulation over time.

Table \ref{tab:max_D} presents the maximum normalized horizontal and vertical diameters computed from the predicted volume fraction fields for five selected samples. In general, the ViViT models generate values that are consistently closer to the expected ones, while the MLP models tend to exhibit larger errors, especially for longer prediction horizons. This behavior is consistent with the previous analyses and highlights the ability of transformer-based architectures to better preserve the droplet dynamics over time.

\newcolumntype{C}[1]{>{\centering\arraybackslash}m{#1}}
\begin{table}[htpb]
\centering
\caption{Maximum normalized horizontal and vertical diameters for five selected test samples. Green cells indicate the best agreement with the expected values, while red cells indicate the worst agreement.}
\begin{tabular}{|m{2.8cm}|
                C{1.0cm}|C{1.2cm}|
                C{1.1cm}|C{1.2cm}|
                C{1.5cm}|C{1.5cm}|
                C{1.6cm}|}
\hline
\multirow{3}{*}{\parbox[c]{2.8cm}{\centering\textbf{Regime and}\\ \textbf{Parameters}\\$(\mathbf{Re},\mathbf{We},\boldsymbol{\beta},\mathbf{Wi})$}}
& \multirow{3}{*}{\parbox[c]{1.0cm}{\centering\textbf{MLP}\\\textbf{50$\rightarrow$1}}}
& \multirow{3}{*}{\parbox[c]{1.2cm}{\centering\textbf{ViViT}\\\textbf{50$\rightarrow$1}}}
& \multirow{3}{*}{\parbox[c]{1.1cm}{\centering\textbf{MLP}\\\textbf{50$\rightarrow$50}}}
& \multirow{3}{*}{\parbox[c]{1.2cm}{\centering\textbf{ViViT}\\\textbf{50$\rightarrow$50}}}
& \multirow{3}{*}{\parbox[c]{1.5cm}{\centering\textbf{MLP}\\\textbf{100$\rightarrow$100}}}
& \multirow{3}{*}{\parbox[c]{1.5cm}{\centering\textbf{ViViT}\\\textbf{100$\rightarrow$100}}}
& \multirow{3}{*}{\parbox[c]{1.6cm}{\centering\textbf{Expected}}} \\

& & & & & & & \\

& & & & & & & \\
\hline

\multicolumn{8}{|c|}{$\mathbf{max(D_H)/D_H(0)}$} \\
\hline

\parbox[c]{2.6cm}{\centering \vspace{2mm} Bouncing \\ $(2.5,0.5,0.9,3)$} & \cellcolor{green!25} 1.1935 & \cellcolor{green!25} 1.1935 & 1.6774 & \cellcolor{green!25} 1.1935 & \cellcolor{red!25} 2.0323 & \cellcolor{green!25} 1.1935 & 1.1935 \\

\parbox[c]{2.6cm}{\centering \vspace{2mm} Bouncing \\ $(51.67,0.5,0.1,3)$} & \cellcolor{green!25} 1.2581 & \cellcolor{green!25} 1.2581 & 1.6774 & \cellcolor{green!25} 1.2581 & \cellcolor{red!25} 2.0323 & \cellcolor{green!25} 1.2581 & 1.2581 \\

\parbox[c]{2.6cm}{\centering \vspace{2mm} Bouncing \\ $(117.22,0.5,0.9,3)$} & \cellcolor{green!25} 1.2581 & \cellcolor{green!25} 1.2581 & 1.6774 & \cellcolor{green!25} 1.2581 & \cellcolor{red!25} 2.0323 & \cellcolor{green!25} 1.2581 & 1.2581 \\

\parbox[c]{2.6cm}{\centering \vspace{2mm} Spreading \\ $(2.5,4.89,0.9,5)$} & 1.2581 & 1.5161 & 1.6774 & \cellcolor{green!25} 1.4516 & \cellcolor{red!25} 2.0323 & \cellcolor{green!25} 1.4516 & 1.3871 \\

\parbox[c]{2.6cm}{\centering \vspace{2mm} Spreading \\ $(100.83,4.89,0.5,3)$} & \cellcolor{green!25} 2.0323 & \cellcolor{green!25} 2.0323 & \cellcolor{red!25} 1.7097 & \cellcolor{green!25} 2.0323 & \cellcolor{green!25} 2.0323 & \cellcolor{green!25} 2.0323 & 2.0323 \\

\hline
\multicolumn{8}{|c|}{$\mathbf{max(D_V)/D_V(0)}$} \\
\hline

\parbox[c]{2.6cm}{\centering \vspace{2mm} Bouncing \\ $(2.5,0.5,0.9,3)$} & 1.0968 & 1.1613 & \cellcolor{red!25} 1.4194 & \cellcolor{green!25} 1.1290 & 1.1875 & \cellcolor{green!25} 1.1290 & 1.1290 \\

\parbox[c]{2.6cm}{\centering \vspace{2mm} Bouncing \\ $(51.67,0.5,0.1,3)$} & 1.1935 & \cellcolor{green!25} 1.2258 & \cellcolor{red!25} 1.4194 & \cellcolor{green!25} 1.2258 & 1.1875 & 1.1935 & 1.2258 \\

\parbox[c]{2.6cm}{\centering \vspace{2mm} Bouncing \\ $(117.22,0.5,0.9,3)$} & \cellcolor{green!25} 1.1935 & 1.2258 & \cellcolor{red!25} 1.4194 & \cellcolor{green!25} 1.1935 & 1.1875 & \cellcolor{green!25} 1.1935 & 1.1935 \\

\parbox[c]{2.6cm}{\centering \vspace{2mm} Spreading \\ $(2.5,4.89,0.9,5)$} & 1.0968 & \cellcolor{green!25} 1.0000 & \cellcolor{red!25} 1.4194 & \cellcolor{green!25} 1.0000 & 1.1875 & \cellcolor{green!25} 1.0000 & 1.0000 \\

\parbox[c]{2.6cm}{\centering \vspace{2mm} Spreading \\ $(100.83,4.89,0.5,3)$} & \cellcolor{green!25} 1.0000 & \cellcolor{green!25} 1.0000 & \cellcolor{red!25} 1.4194 & \cellcolor{green!25} 1.0000 & 1.1875 & \cellcolor{green!25} 1.0000 & 1.0000 \\

\hline
\end{tabular}
\label{tab:max_D}
\end{table}

\section{Conclusion}

The impact of viscoelastic droplets on solid surfaces involves complex interactions between inertia, viscosity, surface tension and elasticity effects. Unlike Newtonian fluids, in which the dynamics are mainly characterized by the Reynolds number $Re$ and the Weber number $We$, viscoelastic droplets require two additional parameters to account for elastic effects, namely the solvent viscosity ratio $\beta$ and the Weissenberg number $Wi$. Different combinations of these parameters result in distinct impact regimes, including bouncing and spreading. Furthermore, full numerical exploration of a large parametric space can become too computationally expensive, especially when considering fine computational meshes.

In this context, machine learning approaches can be used as promising alternatives to accelerate simulations. One of the main difficulties in capturing such dynamics, however, is preserving physical consistency while mitigating error accumulation over time. For this, a method capable of capturing long-range spatial and temporal dependencies is essential. To investigate this, we compared the performance of a simple Multilayer Perceptron (MLP), which is not specifically designed to model spatial and temporal dependencies, with that of a Video Vision Transformer (ViViT) architecture for predicting the temporal evolution of viscoelastic droplets from volume fraction fields. Using only the initial 10\% to 20\% of a numerical simulation as a temporal context input, the models were employed to predict the remaining 80\% to 90\% of the full simulation, resulting in computational time savings of up to 90\%. Additionally, since the proposed framework relies exclusively on volume fraction fields, which are not only available in numerical simulations but can also be extracted from experimental videos, it could be trained or fine-tuned using experimental data, potentially improving the physical consistency and predictive accuracy of the learned dynamics.

The results demonstrated that, although the MLP achieved acceptable performance for one step ahead predictions, its structural limitations became evident as the prediction horizon increased, leading to significant degradation in both structural accuracy and physical consistency. In contrast, the ViViT architecture remained stable across all evaluated scenarios, maintaining high predictive accuracy, enabling more computationally efficient inference when using larger windows and better preserving physical quantities such as geometric properties and contact time.

The qualitative and quantitative analyses showed that ViViT was able to correctly reproduce the impact dynamics associated with the different regimes present in the dataset (spreading and bouncing), while also providing more reliable predictions for longer horizons. These findings demonstrate the potential of transformer-based architectures to model complex droplet impact dynamics characterized by strong nonlinearities and large variations in Reynolds number, Weber number, solvent viscosity ratio and Weissenberg number.

Finally, a user-friendly interface was developed to facilitate the use of the proposed framework. Together with the public availability of the source code, this is intended to encourage interested researchers to reproduce the results presented in this work and explore the proposed methodology in different scenarios, thus contributing to the further development of data-driven approaches for fluid dynamics applications.

\section*{Code availability} \label{sec:code}

The source code developed in this work has been made publicly available. The entire application was implemented in Python and includes a graphical user interface (GUI), allowing users to access its functionalities without requiring extensive programming knowledge or direct interaction with the source code. The implementation can be accessed through the following GitHub repository:

\noindent\textbf{GitHub Repository:} \url{https://github.com/DiegoalexG/FlowML-Toolkit.git}

Any updates and future functionalities will be maintained in the repository.

\section*{Acknowledgements}

This study was financed, in part, by the São Paulo Research Foundation (FAPESP), Brasil, Process Number 2024/15904-7 and Process Number 2024/04769-1, and the National Council for Scientific and Technological Development (CNPq), grant 307228/2023-1. The authors also acknowledge the
Numerical Simulation and AI Laboratory at FCT/UNESP for their support with cluster resources.

\bibliographystyle{alpha}
\bibliography{biblioteca}

@article{Jingzu2023,
	doi = {10.1088/2632-2153/acc727},
	url = {https://dx.doi.org/10.1088/2632-2153/acc727},
	year = {2023},
	publisher = {IOP Publishing},
	volume = {4},
	pages = {025002},
	author = {Jingzu Yee and Daichi Igarashi and Shun Miyatake and Yoshiyuki Tagawa},
	title = {Prediction of the morphological evolution of a splashing drop using an encoder–decoder},
	journal = {Machine Learning: Science and Technology}
}

@article{Hossein2023,
	doi = {doi:10.1515/cppm-2023-0026},
	url = {https://doi.org/10.1515/cppm-2023-0026},
	year = {2023},
	publisher = {De Gruyter},
	author = {Hossein Hassanzadeh and Saptarshi Joshi and Seyed Mohammad Taghavi},
	title = {Predicting buoyant jet characteristics: a machine learning approach},
	journal = {Chemical Product and Process Modeling}
}

@article{Oishi2024,
	doi = {10.1098/rsos.240995},
	url = {https://doi.org/10.1098/rsos.240995},
	year = {2024},
	publisher = {The Royal Society},
	volume = {11},
	pages = {240995},
	author = {Oishi, Cassio M and Kaptanoglu, Alan A and Kutz, J Nathan and Brunton, Steven L},
	title = {Nonlinear parametric models of viscoelastic fluid flows},
	journal = {Royal Society Open Science}
}

@article{Omata2019,
	doi = {10.1063/1.5067313},
	url = {https://doi.org/10.1063/1.5067313},
	year = {2019},
	publisher = {AIP Publishing},
	volume = {9},
	number = {1},
	author = {Omata, Noriyasu and Shirayama, Susumu},
	title = {A novel method of low-dimensional representation for temporal behavior of flow fields using deep autoencoder},
	journal = {Aip Advances}
}

@article{Lorstad2004,
	doi = {https://doi.org/10.1002/fld.746},
	url = {https://onlinelibrary.wiley.com/doi/abs/10.1002/fld.746},
	year = {2004},
	publisher = {John Wiley & Sons},
	volume = {46},
	pages = {109-125},
	author = {Daniel Lörstad and Marianne Francois and Wei Shyy and Laszlo Fuchs},
	title = {Assessment of volume of fluid and immersed boundary methods for droplet computations},
	journal = {International Journal for Numerical Methods in Fluids}
}

@article{Zhuan2022,
	doi = {https://doi.org/10.1002/fld.5061},
	url = {https://onlinelibrary.wiley.com/doi/abs/10.1002/fld.5061},
	year = {2022},
	publisher = {John Wiley & Sons},
	volume = {94},
	pages = {443-460},
	author = {Hai-Zhuan Yuan and Qing Liu and Gang Zeng},
	title = {An adaptive mesh refinement-multiphase lattice Boltzmann flux solver for three-dimensional simulation of droplet collision},
	journal = {International Journal for Numerical Methods in Fluids}
}

@article{Hua2015,
	doi = {https://doi.org/10.1002/fld.3997},
	url = {https://onlinelibrary.wiley.com/doi/abs/10.1002/fld.3997},
	year = {2015},
	publisher = {John Wiley & Sons},
	volume = {77},
	pages = {544-570},
	author = {Hua Tan and Erik Torniainen and David P. Markel and Robert N K Browning},
	title = {Numerical simulation of droplet ejection of thermal inkjet printheads},
	journal = {International Journal for Numerical Methods in Fluids}
}

@misc{Alexey2021,
    year = {2021},
    author = {Alexey Dosovitskiy and Lucas Beyer and Alexander Kolesnikov and Dirk Weissenborn and Xiaohua Zhai and Thomas Unterthiner and Mostafa Dehghani and Matthias Minderer and Georg Heigold and Sylvain Gelly and Jakob Uszkoreit and Neil Houlsby},
    title = {An Image is Worth 16x16 Words: Transformers for Image Recognition at Scale},
    eprint={2010.11929},
    archivePrefix={arXiv},
    primaryClass={cs.CV}
}

@inproceedings{Arnab2021,
  title={Vivit: A video vision transformer},
  author={Arnab, Anurag and Dehghani, Mostafa and Heigold, Georg and Sun, Chen and Lu{\v{c}}i{\'c}, Mario and Schmid, Cordelia},
  booktitle={Proceedings of the IEEE/CVF international conference on computer vision},
  pages={6836--6846},
  year={2021}
}

@inproceedings{Vaswani2017,
    url = {https://proceedings.neurips.cc/paper_files/paper/2017/file/3f5ee243547dee91fbd053c1c4a845aa-Paper.pdf},
    year = {2017},
    publisher = {Curran Associates, Inc.},
    volume = {30},
    author = {Vaswani, Ashish and Shazeer, Noam and Parmar, Niki and Uszkoreit, Jakob and Jones, Llion and Gomez, Aidan N and Kaiser, Lukasz and Polosukhin, Illia},
    booktitle = {Advances in Neural Information Processing Systems},
    title = {Attention is All you Need}
}

@article{Jeremiah1992,
  title={A continuum method for modeling surface tension},
  author={Jeremiah U Brackbill and Douglas B Kothe and Charles Zemach},
  journal={Journal of computational physics},
  volume={100},
  number={2},
  pages={335--354},
  year={1992},
  publisher={Elsevier}
}

@misc{Popinet2013,
  author       = {Stéphane Popinet and collaborators},
  title        = {Basilisk C},
  year         = {2013--2025},
  howpublished = {\url{https://basilisk.fr}}
}

@article{Raanan2005,
  title={Time-dependent simulation of viscoelastic flows at high Weissenberg number using the log-conformation representation},
  author={Raanan Fattal and Raz Kupferman},
  journal={Journal of Non-Newtonian Fluid Mechanics},
  volume={126},
  number={1},
  pages={23--37},
  year={2005},
  publisher={Elsevier}
}

@inproceedings{Takuya2019,
  title={Optuna: A next-generation hyperparameter optimization framework},
  author={Takuya Akiba and Shotaro Sano and Toshihiko Yanase and Takeru Ohta and Masanori Koyama},
  booktitle={Proceedings of the 25th ACM SIGKDD international conference on knowledge discovery \& data mining},
  pages={2623--2631},
  year={2019}
}

@article{Detlef2022,
	doi = {10.1146/annurev-fluid-022321-114001},
	url = {https://doi.org/10.1146/annurev-fluid-022321-114001},
	year = {2022},
	publisher = {Annual Review},
	volume = {54},
	pages = {349-382},
	author = {Detlef Lohse},
	title = {Fundamental Fluid Dynamics Challenges in Inkjet Printing},
	journal = {Annual Review of Fluid Mechanics}
}

@article{Sanjairaj2018,
  title={3D bioprinting of tissues and organs for regenerative medicine},
  author={Sanjairaj Vijayavenkataraman and Wei-Cheng Yan and Wen Feng Lu and Chi-Hwa Wang and Jerry Ying Hsi Fuh},
  journal={Advanced drug delivery reviews},
  volume={132},
  pages={296--332},
  year={2018},
  publisher={Elsevier}
}

@article{Smith2018,
  title={Influence of the impact energy on the pattern of blood drip stains},
  author={Smith, FR and Nicloux, Celine and Brutin, David},
  journal={Physical Review Fluids},
  volume={3},
  number={1},
  pages={013601},
  year={2018},
  publisher={APS}
}

@article{Bolleddula2010,
  title={Impact of a heterogeneous liquid droplet on a dry surface: Application to the pharmaceutical industry},
  author={Bolleddula, DA and Berchielli, A and Aliseda, A},
  journal={Advances in colloid and interface science},
  volume={159},
  number={2},
  pages={144--159},
  year={2010},
  publisher={Elsevier}
}

@article{Gao2025,
  title={A review on thermo-fluidic study of droplet impact in spray cooling},
  author={Gao, Xuan and Li, Yuhang and Xia, Yakang and Li, Haiwang},
  journal={Heat Transfer Research},
  volume={56},
  number={1},
  year={2025},
  publisher={Begel House Inc.}
}

@article{Andrade2013,
  title={Drop impact behavior on food using spray coating: Fundamentals and applications},
  author={Andrade, R and Skurtys, O and Osorio, F},
  journal={Food research international},
  volume={54},
  number={1},
  pages={397--405},
  year={2013},
  publisher={Elsevier}
}

@article{Rein1993,
  title={Phenomena of liquid drop impact on solid and liquid surfaces},
  author={Martin Rein},
  journal={Fluid dynamics research},
  volume={12},
  number={2},
  pages={61--93},
  year={1993},
  publisher={Elsevier}
}

@article{Christophe2004,
  title={Maximal deformation of an impacting drop},
  author={Christophe Clanet and C{\'e}dric B{\'e}guin and Denis Richard and David Qu{\'e}r{\'e}},
  journal={Journal of Fluid Mechanics},
  volume={517},
  pages={199--208},
  year={2004},
  publisher={Cambridge University Press}
}

@article{Li2009,
  title={Drop impact of yield-stress fluids},
  author={Li-Hua Luu and Yo{\"e}l Forterre},
  journal={Journal of Fluid Mechanics},
  volume={632},
  pages={301--327},
  year={2009},
  publisher={Cambridge University Press}
}

@article{Cassio2019,
  title={Normal and oblique drop impact of yield stress fluids with thixotropic effects},
  author={Cassio M Oishi and Roney L Thompson and Fernando P Martins},
  journal={Journal of Fluid Mechanics},
  volume={876},
  pages={642--679},
  year={2019},
  publisher={Cambridge University Press}
}

@article{Kindness2024,
  title={The role of viscoplastic drop shape in impact},
  author={Kindness Isukwem and Julie Godefroid and C{\'e}cile Monteux and David Bouttes and Romain Castellani and Elie Hachem and Rudy Valette and Anselmo Pereira},
  journal={Journal of Fluid Mechanics},
  volume={978},
  pages={A1},
  year={2024},
  publisher={Cambridge University Press}
}

@article{Siddhartha2007,
	doi = {https://doi.org/10.1016/j.ces.2007.07.036},
	url = {https://www.sciencedirect.com/science/article/pii/S0009250907005386},
	year = {2007},
	publisher = {Elsevier},
	volume = {62},
	pages = {7214-7224},
	author = {Siddhartha F. Lunkad and Vivek V. Buwa and K.D.P. Nigam},
	title = {Numerical simulations of drop impact and spreading on horizontal and inclined surfaces},
	journal = {Chemical Engineering Science}
}

@article{Ilia2009,
	doi = {10.1063/1.3129283},
	url = {https://doi.org/10.1063/1.3129283},
	year = {2009},
	publisher = {AIP Publishing},
	volume = {21},
	pages = {052104},
	author = {Ilia V. Roisman},
	title = {Inertia dominated drop collisions. II. An analytical solution of the Navier–Stokes equations for a spreading viscous film},
	journal = {Physics of Fluids}
}

@article{Nick2014,
	doi = {10.1103/PhysRevApplied.2.044018},
	url = {https://link.aps.org/doi/10.1103/PhysRevApplied.2.044018},
	year = {2014},
	publisher = {American Physical Society},
	volume = {2},
	pages = {044018},
	author = {Nick Laan and Karla G. de Bruin and Denis Bartolo and Christophe Josserand and Daniel Bonn},
	title = {Maximum Diameter of Impacting Liquid Droplets},
	journal = {Phys. Rev. Appl.}
}

@article{Jae2016,
	doi = {10.1021/acs.langmuir.5b04557},
	url = {https://doi.org/10.1021/acs.langmuir.5b04557},
	year = {2016},
	publisher = {ACS Publications},
	volume = {32},
	pages = {1299-1308},
	author = {Jae Bong Lee and Dominique Derome and Robert Guyer and Jan Carmeliet},
	title = {Modeling the Maximum Spreading of Liquid Droplets Impacting Wetting and Nonwetting Surfaces},
	journal = {Langmuir}
}

@article{Shiji2018,
	doi = {https://doi.org/10.1016/j.jcis.2017.12.086},
	url = {https://www.sciencedirect.com/science/article/pii/S0021979717314765},
	year = {2018},
	publisher = {Elsevier},
	volume = {516},
	pages = {86-97},
	author = {Shiji Lin and Binyu Zhao and Song Zou and Jianwei Guo and Zheng Wei and Longquan Chen},
	title = {Impact of viscous droplets on different wettable surfaces: Impact phenomena, the maximum spreading factor, spreading time and post-impact oscillation},
	journal = {Journal of Colloid and Interface Science}
}

@article{Jiguo2023,
	doi = {10.1021/acs.iecr.3c01560},
	url = {https://doi.org/10.1021/acs.iecr.3c01560},
	year = {2023},
	publisher = {ACS Publications},
	volume = {62},
	pages = {15268-15277},
	author = {Jiguo Tang and Shengzhi Yu and Xiaofan Hou and Tianhui Wu and Hongtao Liu},
	title = {Universal Model for Predicting Maximum Spreading of Drop Impact on a Smooth Surface Developed Using Boosting Machine Learning Models},
	journal = {Industrial \& Engineering Chemistry Research}
}

@article{Yu2026,
  title={A machine learning-based approach to predict the outcome of binary droplet collision},
  author={Yu, Weidong and Chang, Shinan},
  journal={Chemical Engineering Science},
  volume={319},
  pages={122349},
  year={2026},
  publisher={Elsevier}
}

@article{Vulf2025,
  title={Machine Learning-Based Classification of Suspension Droplet-Solid Wall Impacts for Control of Droplet Fragmentation},
  author={Vulf, Mikhail and Zharikov, Dmitry and Kolomenskiy, Dmitry and Eskin, Dmitry and Osinenko, Pavel},
  journal={IEEE Access},
  year={2025},
  publisher={IEEE}
}

@article{Yu2023,
  title={Robust fluid motion estimator based on attentional transformer},
  author={Yu, Changdong and Chang, Yongpeng and Liang, Xiao and Fan, Yiwei},
  journal={IEEE Transactions on Instrumentation and Measurement},
  volume={72},
  pages={1--14},
  year={2023},
  publisher={IEEE}
}

@article{Solera2024,
  title={$\beta$-variational autoencoders and transformers for reduced-order modelling of fluid flows},
  author={Solera-Rico, Alberto and Sanmiguel Vila, Carlos and G{\'o}mez-L{\'o}pez, Miguel and Wang, Yuning and Almashjary, Abdulrahman and Dawson, Scott TM and Vinuesa, Ricardo},
  journal={Nature Communications},
  volume={15},
  number={1},
  pages={1361},
  year={2024},
  publisher={Nature Publishing Group UK London}
}

@article{Xu2024,
  title={Self-supervised learning based on transformer for flow reconstruction and prediction},
  author={Xu, Bonan and Zhou, Yuanye and Bian, Xin},
  journal={Physics of Fluids},
  volume={36},
  number={2},
  year={2024},
  publisher={AIP Publishing}
}

@article{Khan2025,
  title={Transformer model learning driven analysis of Casson fluid flow influenced by Hall currents and Darcy--Forchheimer effects},
  author={Khan, Muhammad Fawad and Hu, Gang and Sulaiman, Muhammad},
  journal={Physics of Fluids},
  volume={37},
  number={6},
  year={2025},
  publisher={AIP Publishing}
}

@article{Wang2026,
  title={Reconstruction of time-resolved three-dimensional flow fields based on a multi-domain fusion transformer},
  author={Wang, Dezhi and Song, Yuxuan and Chen, Mo and Cao, Yiming and Xia, Weihao and Yu, Changdong},
  journal={Physics of Fluids},
  volume={38},
  number={1},
  year={2026},
  publisher={AIP Publishing}
}

@article{Kazemi2025,
  title={Reservoir Surrogate Modeling Using U-Net with Vision Transformer and Time Embedding},
  author={Kazemi, Alireza and Esmaeili, Mohammad},
  journal={Processes},
  volume={13},
  number={4},
  pages={958},
  year={2025},
  publisher={MDPI}
}

@article{Temizel2025,
  title={Permeability prediction using vision transformers},
  author={Temizel, Cenk and Odi, Uchenna and Li, Kehao and Liu, Lei and Tutun, Salih and Santos, Javier},
  journal={Mathematical and Computational Applications},
  volume={30},
  number={4},
  pages={71},
  year={2025},
  publisher={MDPI}
}

@article{Kang2023,
  title={A new fluid flow approximation method using a vision transformer and a U-shaped convolutional neural network},
  author={Kang, Hyoeun and Kim, Yongsu and Le, Thi-Thu-Huong and Choi, Changwoo and Hong, Yoonyoung and Hong, Seungdo and Chin, Sim Won and Kim, Howon},
  journal={AIP Advances},
  volume={13},
  number={2},
  year={2023},
  publisher={AIP Publishing}
}

@article{Yalamanchi2026,
  title={A Multimodal Vision Transformer-based Modeling Framework for Prediction of Fluid Flows in Energy Systems},
  author={Yalamanchi, Kiran and Barwey, Shivam and Jarrah, Ibrahim and Pal, Pinaki},
  journal={arXiv preprint arXiv:2604.02483},
  year={2026}
}

@inproceedings{Oliu2018,
  title={Folded recurrent neural networks for future video prediction},
  author={Oliu, Marc and Selva, Javier and Escalera, Sergio},
  booktitle={Proceedings of the European conference on computer vision (ECCV)},
  pages={716--731},
  year={2018}
}

@inproceedings{Tran2015,
  title={Learning spatiotemporal features with 3d convolutional networks},
  author={Tran, Du and Bourdev, Lubomir and Fergus, Rob and Torresani, Lorenzo and Paluri, Manohar},
  booktitle={Proceedings of the IEEE international conference on computer vision},
  pages={4489--4497},
  year={2015}
}

@article{Shi2015,
  title={Convolutional LSTM network: A machine learning approach for precipitation nowcasting},
  author={Shi, Xingjian and Chen, Zhourong and Wang, Hao and Yeung, Dit-Yan and Wong, Wai-Kin and Woo, Wang-chun},
  journal={Advances in neural information processing systems},
  volume={28},
  year={2015}
}

@article{Oishi2019,
  title={Impact of capillary drops of complex fluids on a solid surface},
  author={Oishi, CM and Thompson, RL and Martins, FP},
  journal={Physics of Fluids},
  volume={31},
  number={12},
  year={2019},
  publisher={AIP Publishing}
}

@article{Oishi2012,
  title={Numerical simulation of drop impact and jet buckling problems using the eXtended Pom--Pom model},
  author={Oishi, Cassio Machiaveli and Martins, Fernando Pacanelli and Tom{\'e}, Murilo Francisco and Alves, Manuel Ant{\'o}nio},
  journal={Journal of Non-Newtonian Fluid Mechanics},
  volume={169},
  pages={91--103},
  year={2012},
  publisher={Elsevier}
}

@article{Dreisbach2025,
  title={PINNs4Drops: Video-conditioned physics-informed neural networks for two-phase flow reconstruction},
  author={Dreisbach, Maximilian and Kiyani, Elham and Kriegseis, Jochen and Karniadakis, George and Stroh, Alexander},
  journal={arXiv preprint arXiv:2411.15949},
  year={2025}
}

@article{Bertola2009,
  title={An experimental study of bouncing Leidenfrost drops: comparison between Newtonian and viscoelastic liquids},
  author={Bertola, V},
  journal={International journal of heat and mass transfer},
  volume={52},
  number={7-8},
  pages={1786--1793},
  year={2009},
  publisher={Elsevier}
}

@article{Wang2017,
  title={Impact of viscoelastic droplets},
  author={Wang, Yuli and Do-Quang, Minh and Amberg, Gustav},
  journal={Journal of Non-Newtonian Fluid Mechanics},
  volume={243},
  pages={38--46},
  year={2017},
  publisher={Elsevier}
}

@article{Izbassarov2016,
  title={Effects of viscoelasticity on drop impact and spreading on a solid surface},
  author={Izbassarov, Daulet and Muradoglu, Metin},
  journal={Physical Review Fluids},
  volume={1},
  number={2},
  pages={023302},
  year={2016},
  publisher={APS}
}

@article{Jung2013,
  title={The role of viscoelasticity in drop impact and spreading for inkjet printing of polymer solution on a wettable surface},
  author={Jung, Sungjune and Hoath, Stephen D and Hutchings, Ian M},
  journal={Microfluidics and nanofluidics},
  volume={14},
  number={1},
  pages={163--169},
  year={2013},
  publisher={Springer}
}

@article{Crooks2000,
  title={Influence of fluid elasticity on drops impacting on dry surfaces},
  author={Crooks, Regan and Boger, David V},
  journal={Journal of Rheology},
  volume={44},
  number={4},
  pages={973--996},
  year={2000},
  publisher={The Society of Rheology}
}

@article{Rostami2024,
  title={Spreading of a viscoelastic drop on a solid substrate},
  author={Rostami, Peyman and Fricke, Mathis and Schubotz, Simon and Patel, Himanshu and Azizmalayeri, Reza and Auernhammer, G{\"u}nter K},
  journal={Journal of Fluid Mechanics},
  volume={988},
  pages={A51},
  year={2024},
  publisher={Cambridge University Press}
}

\end{document}